\documentstyle[aps,preprint]{revtex}

\tighten

\newcommand{\LSP}{\tilde{\chi}^0_1}
\newcommand{\ben}{\begin{enumerate}}
\newcommand{\een}{\end{enumerate}}
\newcommand{\beq}{\begin{equation}}
\newcommand{\eeq}{\end{equation}}

\newcommand{\slepton}{\tilde{l}}
\newcommand{\selectron}{\tilde{e}}
\newcommand{\smuon}{\tilde{\mu}}

\newcommand{\neutralinoone}{\tilde{\chi}^0_1}
\newcommand{\neutralinotwo}{\tilde{\chi}^0_2}
\newcommand{\charginoone}{\tilde{\chi}^{\pm}_1}

\begin{document}

\draft
\pagestyle{empty}

\preprint{
\noindent
\begin{minipage}[t]{3in}
\begin{flushleft}
\today \\
\end{flushleft}
\end{minipage}
\hfill
\begin{minipage}[t]{3in}
\begin{flushright}
LBNL--40147 \\
UCB--PTH--97/16 \\
FERMILAB-PUB-97/078-T \\
hep-ph/9704205 \\
April 1997
\end{flushright}
\end{minipage}
}

\title{$\bbox{CP}$ Violation from Slepton Oscillations \\ 
at the LHC and NLC
\thanks{This work was supported in part by the Director, Office of
Energy Research, Office of High Energy and Nuclear Physics, Division of
High Energy Physics of the U.S. Department of Energy under Contracts
DE--AC03--76SF00098 and DE--AC02--76CH03000,
and in part by the National Science Foundation under
grant PHY--95--14797.}
} 
\author{Nima Arkani-Hamed, Jonathan L. Feng
\thanks{Research Fellow, Miller Institute for Basic Research in
Science.}, and Lawrence J.~Hall }
\address{Theoretical Physics Group, LBNL and Department of Physics\\ 
University of California, Berkeley, California 94720 }

\author{Hsin-Chia Cheng}
\address{Fermi National Accelerator Laboratory \\
P.~O.~Box 500, Batavia, Illinois 60510}

\maketitle

\begin{abstract}
In supersymmetric theories the charged sleptons of different generations
may oscillate amongst themselves while they decay. In the case of three
generation oscillations, superpartner production at the LHC and NLC may
lead to an observable $CP$-violating signal $N(e^+\mu^-) - N(\mu^+e^-)$.
This signal is proportional to a $CP$-violating invariant of the slepton
mass matrix, $\widetilde{J}$, which is not constrained by searches for
the electric dipole moment of the electron. The sensitivity of the LHC
and NLC to this signal is highly dependent on superpartner masses, but
$\widetilde{J}$ may be probed to a level of $10^{-3}$. Observation of
the $CP$-violating signal would imply a definite structure for the
slepton mass matrices and have strong implications for models of flavor
and SUSY breaking.
\end{abstract}

\pagestyle{plain}

\section{Introduction}

Supersymmetry (SUSY) may provide a solution to the gauge hierarchy problem
and is one of the most attractive candidates for physics beyond the
Standard Model (SM).  If SUSY is discovered, the next experimental task
will be to determine the SUSY parameters as accurately as possible. The
precision measurements of $\sin^2\theta_W$ at LEP and SLD have already
provided us with indirect evidence for supersymmetric unification at
high energy scales. A precise knowledge of superpartner properties could
further test the idea of supersymmetric unification, and may reveal
additional aspects of the theory at high energies, perhaps even up to
the Planck scale.

With so much at stake, much work has been done on the feasibility of
determining SUSY parameters at future colliders\cite{SnowmassSUSY}.
These studies typically exploit precision measurements of kinematic
distributions and cross sections.  At the Large Hadron Collider
(LHC)\cite{LHC}, where many, if not all, of the superpartners are likely
to be produced, recent studies in the minimal supergravity framework
have demonstrated the possibility of precise determinations of some
superpartner mass differences and branching
fractions\cite{SnowmassLHC,Hinchliffe}.  At the proposed Next Linear
Collider (NLC)\cite{NLC}, numerous studies of the
gaugino/Higgsino\cite{JLC,NLCgaugino},
slepton\cite{JLC,BV,NLCslepton,Bartl}, squark\cite{Bartl,NLCsquark}, and
Higgs\cite{NLCHiggs} sectors have shown that if superpartners are
kinematically accessible, highly model-independent measurements of their
masses and properties may be made, and underlying SUSY parameters may be
determined\cite{SnowmassNLC,MP}.

A complementary approach, however, is to look for SUSY-mediated rare
phenomena that violate some of the (approximate) conservation laws of
the SM.  The discovery of these phenomena would be tremendously
exciting, providing essential information about the structure of the
theory that could not be obtained from the studies mentioned above.  In
addition, we will see that these phenomena may be highly sensitive
probes of superpartner mass patterns.  This approach is therefore
particularly well-suited to hadron colliders, where precision
spectroscopy may be very difficult.

In this paper, we explore the phenomenon of $CP$ violation arising from
SUSY-mediated violations of lepton flavor conservation.  Lepton flavor,
although conserved in the SM, is typically violated in any
supersymmetric extension of the SM, since the scalar partners of the
leptons must be given mass, and the scalar mass matrices are generally
not diagonal in the same basis as the fermion masses.  For instance, in
the superfield basis where the lepton Yukawa coupling ${\lambda}_E$
is diagonal, the left- and right-handed slepton masses may be written as

\begin{equation}
\tilde{e}^*_{L\alpha} {m^2_L}_{\alpha\beta} \tilde{e}_{L\beta} 
+\tilde{e}^*_{R\alpha} {m^2_R}_{\alpha\beta} \tilde{e}_{R\beta} \ ,
\end{equation}
where $\alpha$ and $\beta$ are generational indices. The scalar mass
matrices ${m}^2_{L,R}$ are diagonalized by unitary matrices
$W_{L,R}$ through

\begin{equation}
{m}^2_{L,R} = W^{\dagger}_{L,R} {{m}_D^2}_{L,R} W_{L,R} \ ,
\end{equation}
where ${{m}_D^2}_{L,R}$ are diagonal mass matrices.  Together with
their analogues in the quark sector, the $W$ matrices are new flavor
mixing matrices analogous to the CKM matrix. If we work in the mass
eigenstate basis for all fields, the $W$ matrices appear in neutralino
and chargino vertices.  For the neutralinos, these vertices are given by
the interactions

\begin{equation}
\label{convention}
\tilde{e}_{Li} {W_L^*}_{i\alpha} \overline{e_{L\alpha}} \tilde{\chi}^0 
+\tilde{e}^*_{Li} {W_L}_{i\alpha} \overline{{\tilde{\chi}}^0} 
e_{L\alpha} 
+\tilde{e}_{Ri} {W_R^*}_{i\alpha} \overline{e_{R\alpha}} \tilde{\chi}^0 
+\tilde{e}^*_{Ri} {W_R}_{i\alpha} \overline{{\tilde{\chi}}^0} 
e_{R\alpha} \ ,
\end{equation}
where the Latin and Greek subscripts are generational indices for
scalars and fermions, respectively.  

Non-trivial $W$ matrices violate lepton flavor conservation and may give
sizeable contributions to rates for rare processes such as $\mu \to e
\gamma$.  For nearly degenerate sleptons, however, these contributions
are suppressed by $\Delta m/m$ through the superGIM mechanism; in the
limit of exact degeneracy, the $W$ matrices can be rotated to the unit
matrix, and the gaugino interactions conserve lepton flavor.  In fact,
if the mixing between the first two generation sleptons is not very
small, a high degree of degeneracy is {\it required} in order to evade
bounds from $\mu \to e \gamma$.  This is an example of the
supersymmetric flavor-changing problem.

In Ref.~\cite{lfv} we considered high energy probes of the $W$ matrices
through the flavor-violating decays of sleptons at the NLC.  This
flavor-violating signal is only suppressed by the superGIM mechanism
when $\Delta m \alt \Gamma$, and the suppression factor is $\Delta
m/\Gamma$, where $\Gamma$ is the slepton decay width.  As $\Gamma / m$
is typically $10^{-2} - 10^{-3}$, there is a large range of parameters
with $\Gamma \alt \Delta m \alt m$ in which the rates for rare low
energy processes, such as $\mu \to e \gamma$, are suppressed, but the
high energy collider signal is not.  We considered the case of
$\tilde{e}_R - \tilde{\mu}_R$ mixing, and found that the direct collider
probe of $W$ was considerably more powerful than current bounds from
$\mu \to e \gamma$. In this paper, we will extend this analysis by
considering the potential of colliders to probe the phases of the $W$
matrices through $CP$-violating asymmetries in lepton flavor violation.
We will consider both the LHC and the NLC, and will find that both
colliders may be sensitive to small $CP$-violating invariants.

In addition, we will find that any observation of a $CP$ asymmetry
implies a particular structure for the slepton mass matrices, shedding
light on the connection between SUSY breaking and the origin of flavor.
It is well-known that the supersymmetric flavor-changing problem can be
elegantly solved. The two dominant prototypical schemes may be
characterized as follows:
\begin{itemize}
\item
SUSY breaking is mediated to squarks and sleptons by flavor-blind gauge
interactions at a scale well below the Planck scale. In this case the
sfermions of a given charge are degenerate, and the $W$ matrices can be
rotated to the unit matrix. There are no interesting flavor signals in
this scheme, other than the scalar spectrum itself.
\item
Flavor physics of both fermion and scalar sectors is controlled by a
spontaneously broken flavor symmetry. In this case the experimental
signatures of the supersymmetric flavor sector depend on the flavor
symmetry group.  Since the flavor symmetry controls the quark and lepton
masses and mixings, it is frequently the case that the $W$ matrices have
entries that are comparable to the corresponding CKM matrix elements,
and that the scalars of a given charge of the lightest two generations
are closely degenerate.  It may also be that the large interactions
which give the top quark its mass also give a large splitting between
the scalar of the third generation and the other two scalars. In this
case there are many exciting prospects for discovering supersymmetric
flavor physics in rare processes \cite{BHR}. 
\end{itemize}
In the simplest versions of both these cases, the $CP$-violating signals
discussed in this paper will typically be too small to be detected.
However, extra interactions may be present in a more general theory and
there is a general framework in which these $CP$-violating signals are
large:
\begin{itemize}
\item
The dynamics that generates the slepton masses --- whether from
supergravity, gauge, or other interactions --- is flavor-blind, leading
to degenerate scalars. However, the presence of flavor-violating scalar
mass perturbations can lead, via degenerate perturbation theory, to
large mixing angles in $W$ and small scalar non-degeneracies, precisely
the features that lead (as we will show) to large $CP$ asymmetries at
colliders.
\end{itemize}

The outline for the rest of the paper is as follows. In
Sec.~\ref{sec:rephase} we give a rephase-invariant description of $CP$
violation in the slepton mass matrices. In Sec.~\ref{sec:formalism} we
compute the $CP$ asymmetries in lepton flavor-violating events at
colliders in two different ways, first directly from field theory and
then using the more transparent but also more heuristic language of
flavor oscillations.  In Sec.~\ref{sec:scenarios} we determine the
conditions for observable $CP$ violation.  Sec.~\ref{sec:estimate}
contains simple analytic estimates of the $CP$-violating signals at the
LHC and the NLC, and Sec.~\ref{sec:experiment} gives the results of the
numerical calculations and discusses the experimental possibilities for
detecting this signal. In Secs.~\ref{sec:edm} and \ref{sec:mutoegamma},
we consider constraints on the size of the signal coming from the
electron electric dipole moment (EDM) and $\mu \to e \gamma$,
respectively. We discuss the implications of observing a $CP$-violating
signal for models of scalar mass matrices in Sec.~\ref{sec:models} and
draw our conclusions in Sec.~\ref{sec:conclusions}.

\section{A rephase invariant description of slepton $\protect\bbox{CP}$ 
violation}
\label{sec:rephase}

In this section, we count the number of relevant $CP$-violating phases
in the $W_{L,R}$ matrices of Eq.~(\ref{convention}). These phases allow
a variety of $CP$-violating phenomena which we describe in terms of a
well-motivated and complete set of rephase invariant quantities that are
analogous to the Jarlskog invariant of the CKM matrix.  Neglecting the
left-right scalar mass terms, which are suppressed by lepton Yukawa
couplings, slepton mixing may be completely described by the matrices
$W_L$ and $W_R$ of Eq.~(\ref{convention}).  There are then two $3\times
3$ matrices with a combined $9+9=18$ degrees of freedom.  Any physical
quantity must be invariant under rephasings of the slepton and lepton
fields that leave the lepton masses real:

\begin{equation}
{W_L}_{i \alpha} \to e^{i \phi_L^i} {W_L}_{i \alpha} 
e^{i \theta^\alpha}\ , \quad 
{W_R}_{i \alpha} \to e^{i \phi_R^i} {W_R}_{i \alpha}
e^{i \theta^\alpha}\ .
\end{equation}
Thus 9 independent phases may be removed.  However, one overall
rephasing has no effect, and so we can remove 8 degrees of freedom,
leaving 10 parameters to describe $W_{L,R}$. Of these, $3+3$ are real
rotations and 4 are $CP$-violating phases.

We now find a complete set of $CP$ violation rephase invariants,
beginning with the invariants that can be built out of only one of the
two $W$ matrices. As in the SM, there is only one
$CP$-violating phase associated with a single $W$ matrix, and the
corresponding rephase invariant is
\begin{equation}
Im \left[W_{i \alpha} W^*_{i \beta} W^*_{j \alpha} W_{j \beta}\right] 
\equiv
\widetilde{J} \sum_{k\gamma} \epsilon_{ijk}\epsilon_{\alpha\beta\gamma} 
\ .
\end{equation}
$\widetilde{J}_{L}$ and $\widetilde{J}_{R}$ are the supersymmetric
analogues to the Jarlskog invariant\cite{Jarlskog}, 
and the anti-symmetric tensors
$\varepsilon_{ijk}$ and $\varepsilon_{\alpha\beta\gamma}$ simply
determine the relative sign.  In the next section, we will see that the
$CP$ asymmetries in left- and right-handed slepton decays at colliders
are directly proportional to $\widetilde{J}_{L,R}$.  The two remaining
phase degrees of freedom must involve both $W_L$ and $W_R$. In fact, the
dominant contributions to the electron EDM $d_e$ from phases in the $W$
matrices are given by one-loop diagrams with internal sleptons and
gauginos involving left-right scalar mass insertions proportional to
$m_{\mu}$ or $m_{\tau}$.\footnote{Additional contributions may arise
from phases in the $A$, $\mu$ and gaugino mass parameters.  For the
following discussion, we set these to zero.  The justification for this,
and the the relevance of these phases to this study, will be discussed
at the end of this section.}
These contributions therefore require non-trivial flavor mixing for both
left- and right-handed sleptons. If there is only 12 mixing, $d_e$ is
proportional to $m_\mu Im [W_{L21}W^*_{L22}W^*_{R21}W_{R22}]$.  On the
other hand, if there is mixing with the third generation but the first
two generation sleptons are exactly degenerate (or if there is only 13
mixing), $d_e$ is proportional to $m_{\tau}
Im[W_{L31}W^*_{L33}W^*_{R31}W_{R33}]$. Since these two $CP$-violating
quantities are clearly independent and physically meaningful, we choose
the remaining two $CP$-violating invariants to be

\begin{equation} 
\widetilde{K}_{12}=Im\left[{W_L}_{21}{W^*_L}_{22}
{W^*_R}_{21}{W_R}_{22}\right]\ ,
\quad
\widetilde{K}_{13}=Im\left[{W_L}_{31}{W^*_L}_{33}
{W^*_R}_{31}{W_R}_{33}\right]\ .
\label{Ktilde}
\end{equation}
Every $CP$-violating quantity will then be a linear combination of the
four independent invariants $\widetilde{J}_{L,R}$ and
$\widetilde{K}_{12,13}$.

It is also instructive to understand the $CP$-violating phases of the
theory in the superfield basis where the lepton masses are diagonal but
the slepton mass matrices may have off-diagonal components. The only
phases are then in the 12, 23, and 31 components of the slepton mass
matrices. All physical quantities should be invariant under lepton
superfield rephasings ${m^2_{L,R}}_{\alpha\beta} \to e^{-i
\theta^{\alpha}} {m^2_{L,R}}_{\alpha\beta} e^{i \theta^{\beta}}$. If we
concentrate on rephase invariants built from a single mass matrix, the
only possibility is $Im[m^2_{12}m^2_{23}m^2_{31}]$.  Note that this
invariant vanishes unless there is full 3 generation mixing.  It is also
easy to see that if two of the slepton mass eigenvalues are degenerate,
this $CP$ invariant vanishes.  For instance, if the first two generation
sleptons are degenerate at $m_0^2$, we have $m^2_{\alpha\beta} =
\sum_{i} W_{i \alpha}^* W_{i \beta} m^2_i = W_{3\alpha}^* W_{3\beta} 
(m_3^2 - m_0^2)$ for $\alpha \neq \beta$, where $m_i$ are the physical
slepton masses, and we have used the unitarity of $W$, which implies
$\sum_i W^*_{i\alpha} W_{i \beta} = 0$ for $\alpha \neq \beta$. Then,
$Im[m^2_{12} m^2_{23} m^2_{31}]=Im[|W_{31} W_{32} W_{33}|^2 (m_3^2 -
m_0^2)^3] =0$. It is possible to show that (see Appendix \ref{sec:deriveJ}) 

\begin{equation}
\label{vanish}
Im\left[m^2_{12}m^2_{23}m^2_{31}\right] 
= \widetilde{J}(m_2^2 - m_1^2)(m_3^2-m_2^2) (m_1^2 - m_3^2) \ .
\end{equation}
The invariants $\widetilde{K}_{12,13}$ are similarly related to
$Im[{m^{2}_L}^*_{12} {m^2_R}_{12}]$ and $Im[{m^{2}_L}^*_{13}
{m^2_R}_{13}]$.  If we only have mixing between the first two
generations, then

\begin{equation}
Im \left[{m^2_L}^*_{12} {m^2_R}_{12} \right] = 
\widetilde{K}_{12} ({m^2_L}_1 - {m^2_L}_2)({m^2_R}_1 - {m^2_R}_2) \ ,
\end{equation}
whereas if the first two generation sleptons are exactly degenerate,

\begin{equation}
Im\left[{m^{2}_L}^*_{13} {m^2_R}_{13}\right]=\widetilde{K}_{13} 
({m^2_L}_3 - {m^2_L}_0) ({m^2_R}_3 - {m^2_R}_0) \ ,
\end{equation}
where ${m^2_L}_0$ and ${m^2_R}_0$ are the common scalar masses of the
first two generation left- and right-handed sleptons.

Although not needed in the rest of this paper, the $CP$-violating phases
in the squark sector may be similarly counted and classified.  Again
treating the left-right scalar mass mixings as perturbations (an
approximation that may be invalid for the third generation squarks), an
analysis similar to the one above finds that there are 10 independent
phases in the squark mass matrices.\footnote{The counting for only 
left-handed quarks and squarks is considered in Ref.~\cite{Nir}.}  
These also have rephase invariant descriptions
analogous to the $\widetilde{J}$ and $\widetilde{K}$ invariants defined
above.  In the squark sector there are analogous mixing
matrices $W_{U_L}$, $W_{U_R}$, $W_{D_L}$, and $W_{D_R}$. Together
with $V_{\rm CKM}$, four of them are independent ($W_{D_L}$ is fixed by
$W_{U_L}$ and the quark and squark masses).  From these four matrices,
we may form 4 $\widetilde{J}$ invariants (including the Jarlskog invariant
of the CKM matrix), and 2 $\widetilde{K}$
invariants from each of the 3 pairs of matrices \{$W_{U_L}$, $W_{U_R}$\},
\{$W_{D_L}$, $W_{D_R}$\}, and \{$W_{U_L}$, $W_{D_L}$\}.  
These 10 quantities form a complete set.

So far we have discussed only the phases in the $W$ mixing matrices.  Of
course, in addition to these phases, there are other possible sources of
supersymmetric $CP$ violation. The usual SUSY $CP$ problem arises from
relative phases between the $A$ and $\mu$ parameters and the gaugino
masses. In this study, we will examine $CP$-violating asymmetries in
lepton flavor-violating events at colliders.  These will be seen to be
direct measures of the $CP$-violating phases in the $W$ matrices, and
are insensitive to phases in the $A$, $\mu$ and gaugino mass parameters
for the following reasons. First, the $\mu$ parameter and gaugino masses
do not violate lepton flavor, and while their phases may cause changes
in quantities like the total production cross sections, they do not
contribute to flavor-violating asymmetries. Second, while non-universal
$A$ parameters do violate lepton flavor and may contribute to the
$CP$-violating collider signals through left-right scalar mass terms,
these terms are suppressed by lepton masses.  Unless the left- and
right-handed sleptons are very nearly degenerate, the induced left-right
mixing is too small to significantly modify the collider $CP$-violating
asymmetry.

In contrast, the phases in the $A$ and $\mu$ terms and the gaugino
masses may give important contributions to $d_e$.  For this reason, if a
non-zero electron EDM is discovered, it may arise from either these
terms or the $W$ matrices, and does not necessarily imply a lower bound
on the collider signals discussed here.  On the other hand, a bound on
$d_e$ does constrain possible contributions from the $W$ matrices (in
the absence of unexpected cancellations), and therefore also places an
upper bound on our collider signals.  In general, the electron EDM is a
complicated function of both the phases in $W$ and the other phases in
the $A$, $\mu$ and gaugino mass parameters.  However, as we do not know
what these other phases are, we calculate the bounds on $W$ phases from
the electron EDM by assuming that the other phases vanish.  The electron
EDM is then proportional to the invariants $\widetilde{K}_{12,13}$ as
noted above.  We will see in Sec.~\ref{sec:edm} that the collider
signals are observable even if the current limit on $d_e$ improves by a
factor of 10.

To summarize, we have found a set of 4 rephase invariants that
characterize $CP$ violation in the SUSY lepton sector: $\widetilde{J}_L$
and $\widetilde{J}_R$ determine the sizes of $CP$-violating signals at
colliders, while $\widetilde{K}_{12}$ and $\widetilde{K}_{13}$ determine
the size of the electron EDM.  We now consider possible collider probes
of $\widetilde{J}_{L,R}$.

\section{Slepton $\protect\bbox{CP}$ asymmetries}
\label{sec:formalism}

In this section, we derive results for $CP$ asymmetries from slepton
production at colliders.  Our signal will be asymmetries between
flavor-violating final states, such as $e^+ \mu^-$ and $\mu^+ e^-$.  We
will require results for single slepton production, the dominant
production mechanism at hadron colliders where, for example, sleptons
are produced in gluino cascades, and also for correlated slepton pair
production, which is most relevant at $e^+e^-$ colliders.  We begin with
a derivation in field theory, in which various subtleties will be noted
and our approximations explicitly stated.  With these approximations,
the results for single slepton production will be familiar from other
contexts, such as $B$ physics, and we will highlight these similarities.
However, the case of pair production is more complicated, and warrants
careful treatment.  We conclude this section by rederiving these results
in the simpler, but more heuristic, language of flavor oscillations.

\subsection{Field theory derivation}
\label{sec:fieldtheory}

We begin with processes involving single slepton production.  Such
processes are most relevant for hadron colliders, where a single slepton
may be produced in a gluino or squark cascade decay.  The general form
for such a process is $f_1 f_2 \to e^+_{\alpha} X \selectron^-_i \to
e^+_{\alpha} X e^-_{\beta} Y$, where the subscripts $i$, $\alpha$ and
$\beta$ are generational indices.\footnote{In fact, for sleptons produced
in decays of Majorana particles, the same final state also arises from
another process, $f_1 f_2 \to e^-_{\beta} X \selectron^+_i \to
e^-_{\beta} X e^+_{\alpha} Y$, which must be included.  A modification
of the following analysis is necessary, but leads to results identical
to those encapsulated in Eqs.~(\ref{sigmasingle}) and (\ref{sigmapair}).
For notational simplicity, we ignore this complication.}  The initial
state partons are denoted by $f_1$ and $f_2$, and $X$ and $Y$ are $m$-
and $n$-body final states, respectively.  In the simplest case, sleptons
decay directly to LSPs, so $n=1$ and $Y=\LSP$.  For nearly massless
$f_1$ and $f_2$,

\begin{eqnarray}
\sigma_{\alpha\beta} &\equiv& 
\sigma (f_1 f_2 \to e^+_{\alpha} X e^-_{\beta} Y) \nonumber \\
&=& \int \frac{(2\pi)^4}{2\hat{s}} \left| {\cal M}_{\alpha\beta} \right|^2
d\Phi_{m+n+2} ( P;p_{e^+_{\alpha}}, p_{X_1},\ldots,p_{X_m},
p_{e^-_{\beta}}, p_{Y_1},\ldots, p_{Y_n}) \ ,
\end{eqnarray}
where $P=p_{f_1} + p_{f_2}$, $\hat{s} = P^2$, and the phase space factor
and its decomposition are

\begin{eqnarray}
\label{phasespace}
d\Phi_l(P; p_1, \ldots, p_l) &\equiv& \delta^4(P-\sum_{i=1}^l p_i )
\prod_{i=1}^l \frac{d^3p_i}{(2\pi)^3 2E_i} \nonumber \\
&=& d\Phi_{j+1} (P; q, p_1, \ldots, p_j) 
d\Phi_{l-j} (q; p_{j+1}, \ldots, p_l) (2\pi)^3 dq^2 \ ,
\end{eqnarray}
where $q^2 = (\sum_{i=j+1}^l p_i)^2$.  The amplitude ${\cal
M}_{\alpha\beta}$ may be written as

\begin{equation}
{\cal M}_{\alpha\beta} = \sum_i {\cal M}_P W_{i\alpha} 
\frac{i}{q^2 - m^2_i +im\Gamma} W_{i\beta}^* {\cal M}_D \ ,
\end{equation}
where the sum is over all slepton generations that may be produced
on-shell.  ${\cal M}_P$ is the amplitude for $f_1f_2 \to e^+ X
\selectron^-$ in the absence of flavor violation, where $e$ here
represents any one of the available lepton flavors, and ${\cal M}_D$ is
the amplitude for decay $\selectron^- \to e^- Y$, also in the absence of
flavor violation.  We assume that the generational dependence of these
amplitudes, for example, through variations in slepton mass or
dependence on Yukawa couplings, is negligible.  We also neglect
generational differences in the widths $m\Gamma$.

With this amplitude, and substituting the phase space decomposition of
Eq.~(\ref{phasespace}) with $l=m+n+2$ and $j=m+1$,

\begin{equation}
\label{sigmafirst}
\sigma_{\alpha\beta} = \int \frac{(2\pi)^4}{2\hat{s}} 
\left| {\cal M}_P \right|^2 d\Phi_{m+2} (2\pi)^3 
\left| {\cal M}_D \right|^2 d\Phi_{n+1} 
\sum_{ij} W_{i\alpha} W^*_{i\beta} W_{j\alpha}^* W_{j\beta} A_{ij}(q^2)
dq^2 \ ,
\end{equation}
where  

\begin{eqnarray}
A_{ij}(q^2) 
&=&\frac{i}{q^2 - m_i^2 + im\Gamma} \frac{-i}{q^2 - m_j^2 - im\Gamma}
\nonumber \\
&=&\frac{1}{q^2 - \bar{m}_{ij}^2 - \frac{1}{2} \Delta m_{ij}^2+im\Gamma} 
\frac{1}{q^2 - \bar{m}_{ij}^2 + \frac{1}{2} \Delta m_{ij}^2-im\Gamma}\ ,
\end{eqnarray}
and we have defined $\bar{m}_{ij}^2 = (m_i^2 + m_j^2)/2$ and
$\Delta m_{ij}^2 = (m_i^2 - m_j^2)/2 \approx 2m \Delta m_{ij}$, with
$\Delta m_{ij} = m_i-m_j$.  For $\Delta m_{ij}^2, m\Gamma \ll
\bar{m}_{ij}^2$, we may use the approximation

\begin{eqnarray}
\frac{1}{(z-a+ib)(z+a-ib)}
&=& \frac{1}{2(a-ib)} \left[
\frac{z-a-ib}{(z-a)^2+b^2} - \frac{z+a+ib}{(z+a)^2+b^2} \right] 
\nonumber \\
&\approx& \frac{1}{2(a-ib)} \left[
i\pi \delta (z-a) + i \pi \delta (z+a) \right] \nonumber \\
&\approx& \frac{1}{1+i(a/b)}\frac{\pi}{b}\delta (z) \ 
\end{eqnarray}
to write $A_{ij}(q^2) \approx A_{ij} \frac{\pi}{m\Gamma} \delta (q^2 -
\bar{m}_{ij}^2)$. Substituting this form of $A_{ij}(q^2)$ into 
Eq.~(\ref{sigmafirst}) yields

\begin{eqnarray}
\sigma_{\alpha\beta} &=& 
\int \sum_{ij} W_{i\alpha} W^*_{i\beta} W_{j\alpha}^* W_{j\beta} 
A_{ij} \delta(q^2 - \bar{m}_{ij}^2) dq^2 \nonumber \\ 
&& \times \left[\frac{(2\pi)^4}{2\hat{s}}
\left| {\cal M}_P \right|^2 d\Phi_{m+2} \right] 
\left[ \frac{1}{\Gamma} \frac{(2\pi)^4}{2m} 
\left| {\cal M}_D \right|^2 d\Phi_{n+1} \right] \ ,
\end{eqnarray}
and so the general form for the flavor-violating cross section is

\begin{eqnarray}
\sigma_{\alpha\beta}
&=& S_{\alpha\beta} \sigma_0 \nonumber \\
S_{\alpha\beta} &=&
\sum_{ij} W_{i\alpha} W^*_{i\beta} W_{j\alpha}^* W_{j\beta} A_{ij} 
\nonumber \\
A_{ij} &=& \frac{1}{1+ix_{ij}} \ , 
\label{sigmasingle} 
\end{eqnarray}
where we define

\begin{equation}
x_{ij}\equiv \Delta m_{ij}/\Gamma \ ,
\end{equation}
in analogy to the variables $x_d$ and $x_s$ in $B$ physics.  The cross
section $\sigma_0 = \sigma (f_1 f_2 \to e^+ X \selectron^-)
B(\selectron^- \to e^- Y)$ is the analogous cross section in the absence
of flavor violation, where again $e$ here represents any one of the
available lepton flavors. The sum is to be taken over all slepton
generations that may be produced on-shell.  Note that $S_{\alpha\beta}$
is real, since $A_{ij}=A_{ji}^*$.  If the mass splittings are much
larger than the widths, $x_{ij}$ is large and $A_{ij} \approx 0$ for
$i\neq j$, and so only terms with $i=j$ contribute.  However, for
$x_{ij} \alt 1$, interference terms play an important role.

The analysis above assumes that slepton flavor violation arises from
single slepton production.  We now consider the case of correlated
slepton pair production.  Such production may occur at hadron colliders
when sleptons are created in Drell-Yan production, but typically, these
cross sections are too small to be of interest in this study.  At
$e^+e^-$ colliders, however, the dominant slepton production mechanism
is slepton pair production through $s$-channel photon and $Z$ processes
and $t$-channel neutralino exchange.  These processes are then $e^+e^-
\to \selectron^+_i \selectron^-_j \to e^+_{\alpha} X e^-_{\beta} Y$,
where $X$ and $Y$ are again $m$- and $n$-body final states,
respectively.\footnote{For simplicity, throughout this section, we
consider only $\tilde{l}_L\tilde{l}_L$ or $\tilde{l}_R\tilde{l}_R$
production.  Associated production of $\tilde{l}_L\tilde{l}_R$ is also
possible through $t$-channel neutralino exchange.}  For this process,
the most general amplitude is

\begin{eqnarray}
{\cal M}^{\text{pair}}_{\alpha\beta} &=& \sum_i {\cal M}^s_P 
\frac{i}{p^2 - m^2_i +im\Gamma} W_{i\alpha} {\cal M}_D^+
\frac{i}{q^2 - m^2_i +im\Gamma} W_{i\beta}^* {\cal M}_D^- \nonumber\\
&& + \sum_{jk} {\cal M}^t_P 
W_{j\alpha} \frac{i}{q^2 - m^2_j +im\Gamma} W_{j1}^* {\cal M}_D^+
W_{k1} \frac{i}{q^2 - m^2_k +im\Gamma} W_{k\beta}^*{\cal M}_D^- \ ,
\end{eqnarray}
where ${\cal M}^s_P$ and ${\cal M}^t_P$ are the $s$- and $t$-channel
production amplitudes for $e^+e^- \to \selectron^+ \selectron^-$, and
${\cal M}^+_D$ and ${\cal M}^-_D$ are the decay amplitudes for
$\selectron^+ \to e^+ X$ and $\selectron^- \to e^- Y$.  The cross
section is then

\begin{eqnarray}
\sigma^{\text{pair}}_{\alpha\beta} 
&=& \int \frac{(2\pi)^4}{2\hat{s}} 
\left[ \left| {\cal M}^s_P \right|^2 
\sum_{il} W_{i\alpha} W^*_{i\beta} W_{l\alpha}^* W_{l\beta} 
A_{il}(p^2) A_{il}(q^2) \right. \nonumber\\ 
&& + \left( {\cal M}^s_P {\cal M}^{t*}_P
\sum_{imn} W_{i\alpha} W^*_{i\beta} W_{m\alpha}^* W_{m1} W_{n1}^*
W_{n\beta} A_{im}(p^2) A_{in}(q^2) + \text{h.c.} \right) \nonumber\\ 
&& + \left. \left| {\cal M}^t_P \right|^2 
\sum_{jkmn} W_{j\alpha} W^*_{j1} W_{k1} W_{k\beta}^* 
W_{m\alpha}^* W_{m1} W_{n1}^* W_{n\beta} A_{jm}(p^2) A_{kn}(q^2) \right]
\nonumber\\ 
&& \times dp^2 dq^2 d\Phi_2 
(2\pi)^3 \left| {\cal M}^+_D \right|^2 d\Phi_{m+1} 
(2\pi)^3 \left| {\cal M}^-_D \right|^2 d\Phi_{n+1} \ .
\end{eqnarray}
Letting $\sigma_0^{ss}$, $\sigma_0^{st}$, and $\sigma_0^{tt}$ denote the
products of $B(\selectron^+ \to e^+ X) B(\selectron^- \to e^- Y)$ and
the flavor-conserving ``cross sections'' corresponding to production
amplitudes $\left| {\cal M}^s_P \right|^2$, ${\cal M}^s_P {\cal
M}^{t*}_P$, and $\left| {\cal M}^t_P \right|^2$, respectively, we find

\begin{eqnarray}
\sigma^{\text{pair}}_{\alpha\beta} &=& 
S_{\alpha\beta}^{ss}\sigma_0^{ss} 
+ \left( S_{\alpha\beta}^{st}\sigma_0^{st} + \text{h.c.}\right)
+ S_{\alpha\beta}^{tt}\sigma_0^{tt} \nonumber \\
S_{\alpha\beta}^{ss} &=& \sum_{ij} W_{i\alpha} W^*_{i\beta} 
W_{j\alpha}^* W_{j\beta} A_{ij}^2 \nonumber \\ 
S_{\alpha\beta}^{st} &=& 
\sum_{ijk} W_{i\alpha} W^*_{i\beta} W_{j\alpha}^* W_{j1} W_{k1}^*
W_{k\beta} A_{ij} A_{ik} \nonumber \\ 
S_{\alpha\beta}^{tt} &=& S_{\alpha 1} S_{1\beta} \ .
\label{sigmapair}
\end{eqnarray}
$S_{\alpha\beta}^{tt}$ decomposes into a product of single slepton
factors $S_{\alpha\beta}$, as expected for uncorrelated production from
two vertices. 

The general expressions for $\sigma_{\alpha\beta}$ and
$\sigma^{\text{pair}}_{\alpha\beta}$ given in Eqs.~(\ref{sigmasingle})
and (\ref{sigmapair}) may be applied to a number of interesting
scenarios.  In particular, let us consider the case where

\begin{equation}
W = \left(
\begin{array}{ccc}
 \cos\theta & \sin\theta & 0 \\
-\sin\theta & \cos\theta & 0 \\
0           & 0          & 1 \end{array} \right) \ ,
\end{equation}
and both $\tilde{e}$ and $\tilde{\mu}$ may be produced on-shell.
Substituting this $W$ into Eq.~(\ref{sigmasingle}) and letting

\begin{equation}
\chi_{ij} \equiv \frac{x_{ij}^2}{2(1+x_{ij}^2)} \ ,
\end{equation}
in analogy to the parameters $\chi_d$ and $\chi_s$ in $B$ physics, we
find

\begin{equation}
S_{12} = 4 \chi_{12} \sin^2\theta \cos^2\theta \ ,
\end{equation}
reproducing the 2 generation flavor-violating result given in
Ref.~\cite{lfv}.  The pair production case is more complicated, but the
final results are

\begin{eqnarray}
S^{ss}_{12}&=&4 \chi_{12}(3-4\chi_{12}) \sin^2\theta\cos^2\theta \\ 
Re \left[ S^{st}_{12}\right] &=&
2 \chi_{12}(3-4\chi_{12}) \sin^2\theta\cos^2\theta \\
S^{tt}_{12}&=&S_{11}S_{12} = S_{12}(1-S_{12}) \ .
\end{eqnarray}
These quantities are unchanged if $\tilde{\tau}$ 
pairs may also be
produced.  

Finally, we derive expressions for $CP$-violating asymmetries.  In the
presence of $CP$ violation, the cross sections $\sigma_{\alpha\beta} =
\sigma (f_1 f_2 \to e^+_{\alpha} X e^-_{\beta} Y)$ and
$\sigma_{\beta\alpha} = \sigma (f_1 f_2 \to e^+_{\beta} X e^-_{\alpha}
Y)$ are no longer equal. For the case of single slepton production, it
is easy to show that, following the analysis above,

\begin{equation}
\Delta_{\alpha\beta} \equiv S_{\alpha\beta} - S_{\beta\alpha} =
\frac{\sigma_{\alpha\beta} - \sigma_{\beta\alpha}}{\sigma_0} = 
-4 \sum_{i<j} Im \left[ W_{i\alpha} W^*_{i\beta} W_{j\alpha}^* W_{j\beta} 
\right] Im \left[ A_{ij} \right] \ .
\end{equation}
Again, the sum is over sleptons that may be produced on-shell.  We
therefore reproduce the familiar result that $CP$ violation requires the
presence of at least two amplitudes that differ both in their $CP$-odd
(``weak'') phases and their $CP$-even (``strong'') phases.  The
$W$-dependent part may be written as

\begin{equation}
Im \left[ W_{i\alpha} W^*_{i\beta} W_{j\alpha}^* W_{j\beta} \right] 
= \widetilde{J} \sum_{k\gamma} \varepsilon_{ijk}
\varepsilon_{\alpha\beta\gamma} \equiv \widetilde{J}^{ij}_{\alpha\beta}\ , 
\end{equation}
where $\widetilde{J}$ is the supersymmetric analogue to the Jarlskog
invariant introduced in Sec.~\ref{sec:rephase}.  All single slepton
$CP$-violating quantities are given in terms of $\widetilde{J}$, the
single $CP$-violating invariant that may be formed from one $3\times 3$
slepton mixing matrix.  The $CP$-violating single slepton cross section
asymmetry is then

\begin{equation}
\label{delta}
\Delta_{\alpha\beta} =  
-4 \sum_{i<j} \widetilde{J}^{ij}_{\alpha\beta} Im [A_{ij}] =
4 \sum_{i<j} \widetilde{J}^{ij}_{\alpha\beta} 
\frac{x_{ij}}{1+x_{ij}^2} \ .
\end{equation}
For all $\alpha \ne \beta$, the $\Delta_{\alpha\beta}$ are equal up to a
sign, and so we see that there is only one $CP$-violating quantity, and
the $e\mu$, $\mu\tau$, and $\tau e$ asymmetries are all measurements of
the same underlying $CP$-violating parameter.

For pair production, the asymmetry is

\begin{equation}
\sigma^{\text{pair}}_{\alpha\beta}-\sigma^{\text{pair}}_{\beta\alpha} = 
\Delta_{\alpha\beta}^{ss}\sigma_0^{ss} 
+ \left( \Delta_{\alpha\beta}^{st}\sigma_0^{st} + \text{h.c.}\right)
+ \Delta_{\alpha\beta}^{tt}\sigma_0^{tt} \ ,
\end{equation}
where

\begin{eqnarray}
\Delta_{\alpha\beta}^{ss} &\equiv& 
S_{\alpha\beta}^{ss} - S_{\beta\alpha}^{ss} = 
-4\sum_{i<j} \widetilde{J}^{ij}_{\alpha\beta} Im [ A_{ij}^2 ] \\
\Delta_{\alpha\beta}^{st} &\equiv& 
S_{\alpha\beta}^{st} - S_{\beta\alpha}^{st} = 
2 i \sum_{ijk} Im [ W_{i\alpha} W_{i\beta}^* 
W_{j\alpha}^* W_{j1} W_{k1}^* W_{k\beta} ] A_{ij} A_{ik} \\
\Delta_{\alpha\beta}^{tt} &\equiv& 
S_{\alpha\beta}^{tt} - S_{\beta\alpha}^{tt} = 
S_{\alpha 1}S_{1 \beta} - S_{\beta 1}S_{1 \alpha} =
\Delta_{1\beta} S_{1\alpha} - \Delta_{1\alpha} S_{1\beta} \ .
\end{eqnarray}
The pair asymmetry must also be proportional to $\widetilde{J}$, and it
is clear from the expressions above that $\Delta_{\alpha\beta}^{ss}$ and
$\Delta_{\alpha\beta}^{tt}$ are.  For $\Delta_{\alpha\beta}^{st}$, this
dependence may also be made manifest, as is shown in
Appendix~\ref{sec:stinterference}.

\subsection{Oscillation derivation}

In this section, we show how the above results derived directly from
field theory can be reproduced using the more familiar language of
flavor oscillations. Suppose that at time $t=0$, we produce
$e^+_{\alpha}$ in association with a {\em gauge} eigenstate slepton
$\tilde{e}_{\alpha}^-$.  We denote this initial state by
$|\psi(0)\rangle = |\alpha^-\rangle$.  This state then oscillates, and
the amplitude for the slepton to decay into $e^-_{\beta}$ is

\begin{equation}
A_{\alpha^- \to \beta^-}(t) = \langle \beta^- |e^{-i \bbox{m} t}
e^{-\Gamma t/2} |\alpha^- \rangle = \sum_{i} W_{i \alpha} W^*_{i \beta}
e^{-i m_i t} e^{-\Gamma t/2} \ ,
\end{equation}
where the sum is over all generations.\footnote{The oscillation picture
is applicable only if all the sleptons are nearly degenerate. Otherwise,
the amplitude for producing the initial gauge eigenstate slepton is
ill-defined.  For instance, suppose that the gauge eigenstate is a
superposition of a mass eigenstate beneath production threshold and a
mass eigenstate above threshold.  The amplitude for producing the gauge
eigenstate slepton is clearly not well-defined in this case.}  The
corresponding amplitude for anti-sleptons, $A_{\alpha^+\to\beta^+}(t)$,
is the same with $W \leftrightarrow W^*$. The probability that
$e_{\beta}^-$ is produced between $t$ and $t + dt$ is $\Gamma dt
|A_{\alpha^- \to \beta^-}(t)|^2$, so the total probability of producing
an $e_{\alpha}^+ e_{\beta}^-$ final state is (throughout we use
$\int_{0}^{\infty} \Gamma dt e^{-i(m_i - m_j)t} e^{-\Gamma t} = \Gamma
/(\Gamma + i(m_i - m_j)) \equiv A_{ij}$)

\begin{equation}
P_{\alpha^- \to \beta^-} = 
\int_{0}^{\infty} \Gamma dt |A_{\alpha^- \to \beta^-}(t)|^2 
= \sum_{ij} W_{i \alpha} W^*_{i \beta} W^*_{j\alpha} W_{j \beta} 
A_{ij} \ ,
\end{equation}
which reproduces Eq.~(\ref{sigmasingle}) in the previous section.

Turning now to the case where correlated sleptons are produced from $s$-
and $t$-channel diagrams, the initial state is
\begin{equation}
|\psi(0)\rangle = {\cal M}_P^s \sum_{\gamma}|\gamma^+,\gamma^- \rangle
+ {\cal M}_P^t |1^+, 1^- \rangle \ , 
\end{equation}
where ${\cal M}_P^s$ and ${\cal M}_P^t$ are the $s$- and $t$-channel
amplitudes, respectively.  The amplitude for producing $e^+_{\alpha}$ at
time $t_1$ and $e^-_{\beta}$ at time $t_2$ is

\begin{equation}
A^{\text{pair}}_{\alpha^+ \beta^-}(t_1,t_2) 
= {\cal M}_P^s \sum_{\gamma}
A_{\gamma^+ \to \alpha^+}(t_1) A_{\gamma^- \to \beta^-}(t_2) 
+ {\cal M}_P^t A_{1^+ \to \alpha^+}(t_1) A_{1^- \to \beta^-}(t_2) \ .
\end{equation}
However,

\begin{eqnarray}
\sum_{\gamma} A_{\gamma^+ \to \alpha^+}(t_1) A_{\gamma^- \to 
\beta^-}(t_2)
&=&\sum_{\gamma i j}
W^*_{i \gamma} W_{i \alpha} W_{j \gamma} W^*_{j \beta} e^{-i m_i t_1}
e^{-\Gamma t_1/2} e^{-i m_j t_2} e^{-\Gamma t_2/2} \nonumber \\ 
&=&\sum_{i} W_{i \alpha} W^*_{i \beta} e^{-i m_i t_1} e^{-\Gamma t_1/2}
e^{-i m_i t_2} e^{-\Gamma t_2/2} \ ,
\end{eqnarray}
so

\begin{eqnarray}
A^{\text{pair}}_{\alpha^+ \beta^-}(t_1,t_2) 
&=& {\cal M}_P^s \sum_{i} W_{i \alpha} W^*_{i \beta} e^{-i m_i t_1} 
e^{-\Gamma t_1/2} e^{-i m_i t_2} e^{-\Gamma t_2/2} \nonumber \\
&&+ {\cal M}_P^t A_{1^+ \to \alpha^+}(t_1) A_{1^- \to \beta^-}(t_2) \ .
\end{eqnarray} 
Therefore, the total probability of producing $e^+_{\alpha} e^-_{\beta}$
is

\begin{eqnarray}
P^{\text{pair}}_{\alpha^+\beta^-}&=&
\int_{0}^{\infty} \Gamma dt_1 \Gamma dt_2 
|A^{\text{pair}}_{\alpha^+ \beta^-} (t_1,t_2)|^2  \nonumber \\
&=&
|{\cal M}_P^s|^2 \sum_{ij} W_{i \alpha} W^*_{i \beta} 
W^*_{j \alpha} W_{j \beta} A_{ij}^2 \nonumber \\
&&+ \left( {\cal M}_P^s {\cal M}_P^{t*} 
\sum_{ijk} W_{i \alpha} W^*_{i \beta} W_{j 1} W^*_{j \alpha}
W^*_{k 1} W_{k \beta} A_{ij} A_{ik} + \text{h.c.} \right) \nonumber \\
&&+ |{\cal M}_P^t|^2 P_{1^+ \to \alpha^+} P_{1^- \to \beta^-}  \ ,
\end{eqnarray}
in agreement with Eq.~(\ref{sigmapair}) of the previous section.

\section{Scenarios with Observable $\protect\bbox{CP}$ Violation}
\label{sec:scenarios}

In the previous section, we derived formulas for $CP$-violating cross
sections, which depended on the slepton mixing angles and mass
splittings.  In this section, we examine these expressions and determine
what conditions must be met to yield promising $CP$-violating signals.
We will find that $CP$ violation in the lepton sector may be large in
two scenarios.  These two scenarios will be used in the following
sections to determine the typical reaches in parameter space of collider
experiments.

The $CP$-violating difference in cross sections was derived in
Sec.~\ref{sec:fieldtheory} for both single and correlated pair slepton
production.  Here we will analyze the simpler case of single slepton
production, as the conclusions are based on general arguments that apply
in both cases.  Recall that the $CP$-violating difference in cross
sections in the case of single slepton production is given by

\begin{equation}
\Delta_{\alpha\beta} = \frac{\sigma_{\alpha\beta} -
\sigma_{\beta\alpha}} {\sigma_0} = 4
\sum_{i<j}\widetilde{J}_{\alpha\beta}^{ij} \frac{x_{ij}}{1+x_{ij}^2} \ .
\end{equation}
In the typical case where all three slepton generations are produced,

\begin{equation}
\Delta_{e \mu} = \Delta_{\mu \tau} =  \Delta_{\tau e}
= 4 \widetilde{J} \left( \frac{x_{12}}{1 + x_{12}^2} +
\frac{x_{23}}{1 + x_{23}^2} +\frac{x_{31}}{1 + x_{31}^2} \right) \ .
\label{deltaemu}
\end{equation}
For a significant $CP$-violating signal, both $\widetilde{J}$ and the
kinematic part depending on the mass splittings $x_{ij} \equiv \Delta
m_{ij} / \Gamma$ must be fairly large.

Concentrating first on $\widetilde{J}$, we recall that in the
parametrization

\begin{equation}
W = \left(
\begin{array}{ccc}
c_{12} c_{13} & s_{12}c_{13}& s_{13}e^{-i\delta} \\
-s_{12} c_{23} - c_{12}s_{23}s_{13}e^{i\delta}&
c_{12} c_{23} - s_{12}s_{23}s_{13}e^{i\delta}&s_{23}c_{13} \\
s_{12} s_{23} - c_{12}c_{23}s_{13}e^{i\delta}&
-c_{12} s_{23} - s_{12}c_{23}s_{13}e^{i\delta}&c_{23}c_{13}
\end{array} \right) \ ,
\end{equation}
where $s_{ij} = \sin \theta_{ij}$ and $c_{ij} = \cos
\theta_{ij}$\cite{PDG,Chau}, the supersymmetric Jarlskog invariant is 
$\widetilde{J} = s_{12} s_{13} s_{23} c_{12} c_{13}^2 c_{23} \sin
\delta$, with maximal value $\frac{1}{6\sqrt{3}} \approx 0.096$.  
Large $CP$ violation therefore requires large three generation mixing
and a large $CP$-violating angle.  The $CP$-violating signal for single
slepton production is completely determined by $\widetilde{J}$.
However, to evaluate backgrounds, and to compute the $CP$-violating
signal at the NLC from correlated slepton pair production, we need to
know the individual angles $\theta_{ij}$ of the $W$ matrix, rather than
just the combination $\widetilde{J}$. For the numerical calculations of
the following sections, we choose

\begin{equation}
\theta \equiv \theta_{12} =\theta_{23}=\theta_{13} \ .
\end{equation}
In this parametrization, $\widetilde{J} = \sin^3 \theta \cos^4\theta
\sin \delta$, and for $\sin\theta = \sqrt{\frac{3}{7}}$ ($\sin 2\theta =
\frac{4\sqrt{3}}{7} \approx 0.99$), $\widetilde{J}$ attains its maximal 
value $\frac{48\sqrt{3}}{343\sqrt{7}} \approx 0.092$. This simplifying
assumption therefore captures nearly all of the available range of
$\widetilde{J}$.

Turning now to the kinematic factor, it is clear that it may be large
only if at least one $x_{ij}$ is near 1, since each term is maximal for
$x_{ij}=1$, and drops off for larger and smaller $x_{ij}$.  In addition,
since $x_{12}+x_{23}+x_{31}=0$, the signal is suppressed if any pair of
sleptons is highly degenerate, in accord with Eq.~(\ref{vanish}).  There
are therefore two possible scenarios with large $CP$ violation: either
(I) one $x_{ij}$ is ${\cal O}(1)$ and the other two are much larger than
1, or (II) all three $x_{ij}$ are roughly ${\cal O}(1)$.  As will be
discussed in Sec.~\ref{sec:mutoegamma}, the three sleptons must have
some degree of degeneracy ($x_{ij} \alt 10-100$) to ensure that the
large mixing angles of $W$ do not induce too large a branching ratio for
$\mu \to e \gamma$.  For concreteness, we assume that $m_1 > m_2 > m_3$
and in each scenario we parametrize the mass splittings in terms of a
single variable $x$:

\begin{eqnarray}
\text{Scenario I}: \ x &\equiv& x_{12} \ , \quad 0.1 \le x \le 10 \ ,
\quad x_{23} = 10 \\
\text{Scenario II}: \ x &\equiv& x_{12} = x_{23}\ , \quad 0.1 \le x 
\le 10 \ .
\end{eqnarray}
With these parametrizations, in both Scenarios I and II the signal and
background cross sections depend on only three parameters: $x$,
$\theta$, and $\delta$.

In summary, we can identify the necessary features of the slepton mass
matrices if a $CP$-violating slepton oscillating signal is to be visible
at future colliders:

\begin{enumerate}
\item The $W$ matrices must have large $CP$- and flavor-violating angles.
\item The two sleptons of highest degeneracy should have $\Delta m$ of 
order $\Gamma$, giving $x$ of order unity.
\item All three sleptons must have some degree of degeneracy, so that 
$x_{ij} \alt 10 - 100$.
\end{enumerate}
There conditions may be satisfied in two regions of parameter space,
which we denote Scenarios I and II.

\section{Analytic estimate of the LHC and NLC signals}
\label{sec:estimate}

$CP$ violation in the slepton sector may be detected at colliders by
considering a sample of dilepton events and observing a statistically
significant asymmetry
 
\begin{equation}
\label{asymmetry}
A \equiv \frac{S}{B} = 
\frac{N_{e^+_{\alpha} e^-_{\beta}} - N_{e^+_{\beta} e^-_{\alpha}}}
{N_{e^+_{\alpha} e^-_{\beta}} + N_{e^+_{\beta} e^-_{\alpha}}} \ ,
\end{equation}
where $\alpha$ and $\beta$ are generational indices.  The numerator is
the $CP$-violating signal.  The denominator, the ``background,''
includes all events that pass the cuts, and includes contributions from
the flavor-changing slepton events themselves, as well as other
backgrounds.  To maximize the statistical significance of the asymmetry,
we must isolate a large and pure sample of slepton events.

In the following section, we will present precise calculations of the
signal and backgrounds at the LHC and NLC. In this section, however, we
give rough order-of-magnitude estimates, with a view to gaining a
general understanding of what ranges of flavor parameters may be probed.
The validity of these estimates is justified for certain SUSY parameters
by the detailed analysis in the following section.

For optimal mass splittings, the signal is roughly $S \approx
\sigma_S \varepsilon \widetilde{J} L$.  $L$ is the total 
integrated luminosity, and $\sigma_S$ is the production cross section of
the relevant superparticles --- squarks and gluinos for hadron
colliders, and sleptons for lepton colliders (see below).  The
efficiency $\varepsilon$ is the fraction of such superparticle events
that end up in the final sample of slepton events (for any one of the
available slepton generations) and therefore includes all branching
ratios and kinematic cut efficiencies.  Denoting the background by
$B=\sigma_B L$, we find that a $3\sigma$ signal requiring $S >
3\sqrt{B}$ implies that a $CP$-violating signal may be observed for

\begin{equation}
\label{estimate}
\widetilde{J} \agt \frac{3}{\sigma_S \varepsilon}
\sqrt{\frac{\sigma_B}{L}} \ ,
\end{equation}
where the cross sections are in fb, and $L$ is in fb$^{-1}$. 

At the LHC with $\sqrt{s} = 14$ TeV, the most promising source of
sleptons is in cascade decays $(\tilde{g} \to )\, \tilde{q}_L \to
\neutralinotwo \to \slepton \to \neutralinoone$.  Such cascades may be
prominent when the charginos and neutralinos are gaugino-like, and
$m_{\tilde{l}} < m_{\neutralinotwo}$.  The efficiency is typically
$\varepsilon \sim 1\%$.  The background is dominated by flavor-changing
slepton events and gluino/squark pair production leading to two
leptonically-decaying charginos.  Assuming $L = 100 \text{ fb}^{-1}$ and
$m_{\tilde{g}}, m_{\tilde{q}} \approx 300$ (700) GeV, the  
cross sections of gluino/squark pair production and the background
we may expect are $\sigma_S \sim 1000$ (10) pb and $\sigma_B
\sim 1000$ (10) fb, and from Eq.~(\ref{estimate}) we find that a 
3$\sigma$ signal requires $\widetilde{J} \agt 10^{-3} \, (10^{-2})$.

At the LHC the efficiency $\varepsilon$ is a product of many factors.
Of the total strong production cross section, only events containing
$\tilde{q}_L$ may produce sleptons, as right-handed squarks decay
directly to $q \LSP$.  In contrast, $B(\tilde{q}_L \to q
\tilde{\chi}_2^0 \, (q \tilde{\chi}_1^+)) = 1/3\, (2/3)$, since 
$B(\tilde{q}_L \to q\LSP)$ is suppressed relative to these by
hypercharge couplings.  Decays to charginos are also not useful, as the
lepton flavor-changing information is carried off in
neutrinos.\footnote{For the case that $\tilde{\chi}_1^{\pm}$ can decay
to all three generations of sleptons, the branching fraction for decays
$\tilde{\chi}_1^- \to e^-_\alpha\LSP$ and $\tilde{\chi}_1^+ \to
e^+_\alpha\LSP$ are both proportional to $\sum_\beta S_{\alpha \beta} =
1$, and so do not contribute to a $CP$ asymmetry.} 
Another reduction in efficiency arises
in the branching fraction $B(\tilde{\chi}_2^0 \to e^+_\alpha e^-_\beta
\LSP)$.  Assuming for simplicity that the produced sleptons are
left-handed,

\begin{equation}
B(\tilde{\chi}_2^0 \to e^+_\alpha e^-_\beta \LSP) =
\frac{S_{\alpha \beta}}{2N} \ ,
\end{equation}
where $N=3$ is the number of generations of sleptons to which
$\tilde{\chi}_2^0$ can decay, and the factor of 1/2 accounts for the
fact that we include only the decays to charged sleptons.  Finally,
there is the efficiency of kinematic cuts, which is typically $\agt
10\%$.

At the NLC sleptons are produced directly in pairs.  The signal is
therefore not degraded by branching ratios, and typical efficiencies for
the kinematic cuts are $\varepsilon \approx 30\%$.  For correlated pair
production, the signal is no longer simply proportional to
$\widetilde{J}$.  However, for a rough estimate, we can again apply the
above analysis.  Assuming 200 GeV sleptons are produced at a $\sqrt{s} =
500$ GeV collider with $L=50 \text{ fb}^{-1}$, the signal cross section
is $\sigma_S \sim 300$ fb.  The background is largely flavor-violating
slepton events and so varies with $\widetilde{J}$; in the regions of
$\widetilde{J}$ we can probe, $\sigma_B \sim 10$ fb. A 3$\sigma$ signal
requires $\widetilde{J} \agt 10^{-2}$.  However, for $L=500
\text{ fb}^{-1}$, we expect to be able to probe to the $\widetilde{J}
\sim {\cal O}(10^{-3})$ level.

In this paper, we will concentrate on signals at the LHC and NLC,
because, as seen from the above rough estimates, these machines offer
the possibility of probing $\widetilde{J}$ to the level of $10^{-3}$,
well below the maximum of value of $\frac{1}{6 \sqrt{3}}$. However,
before analyzing these colliders' capabilities in detail, it is
worthwhile to consider whether such signals may be seen at LEP II or the
Tevatron.  At LEP II with $\sqrt{s} = 190$ GeV, for slepton masses of 80
GeV, just above the current bound, $\sigma_S$ may be $\sim 500$ fb,
depending on the neutralino masses.  Systems of cuts for
flavor-conserving slepton events\cite{LEPIIWG} may be adopted for
flavor-violating events\cite{lfv} and typically give values of
$\varepsilon \approx 40\% - 60\%$ and $\sigma_B \approx 10-100$ fb.
Assuming a total integrated luminosity of $500 \text{ pb}^{-1}$, we
found in Ref.~\cite{lfv} that lepton flavor violation could indeed be
observed for large mixing angles.  Substituting these ranges of
quantities into Eq.~(\ref{estimate}), we find the requirement
$\widetilde{J} \agt 0.04 - 0.2$; an observable $CP$-violating signal may
therefore be possible for near-maximal $\widetilde{J}$. At Run II and
future upgrades of the Tevatron with $\sqrt{s} = 2$ TeV, if squark and
gluino masses are in the region of 200 GeV, the total strong SUSY
production cross section is $\sigma_S \sim 10$ pb.  Given $L \sim 2-30
\text{ fb}^{-1}$, the number of signal events is roughly only an order
of magnitude below the number of events at the LHC with 700 GeV gluinos
and squarks.  It is therefore possible that flavor-violating slepton
signals could be observed, and possibly even $CP$-violating asymmetries
could be seen for near-maximal $\widetilde{J}$.  However, such
possibilities depend crucially on many factors, such as the
effectiveness of cuts to isolate the slepton signal from $t\bar{t}$ and
other backgrounds, and will not be explored in detail here.

\section{Experimental possibilities for the LHC and NLC}
\label{sec:experiment}

In this section, we will consider the possibilities for detecting
slepton $CP$ violation in two experimental environments: the LHC at CERN
and the NLC, a proposed high energy linear $e^+e^-$ collider.  The
sensitivity of these two colliders to slepton $CP$ violation is
dependent on many of the underlying SUSY parameters. These parameters
set the masses and branching fractions that determine the number of
sleptons produced.  In addition, they fix the rates and kinematic
distributions of the many background processes, which determine the
extent to which a pure slepton signal may be isolated.  A comprehensive
analysis would require a scan of parameter space and optimization of
cuts for each parameter set and is beyond the scope of this study.

Instead, to gain an understanding of what sensitivities are typically
achievable and what aspects of the sparticle spectrum are most
important, we will limit our analysis by considering a particular 
set of SUSY parameters for each of the two
colliders.  This choice will determine our signal and background rates,
and will also allow us to use systems of cuts that have been developed
previously.  After presenting the results, we will also discuss the
implications for our analysis of variations away from these parameters.

\subsection{LHC} 
\label{LHC}

The LHC is a $pp$ collider with $\sqrt{s} = 14$ TeV and luminosity
${\cal L} \sim 10$--100 fb$^{-1}$/yr.  At hadron colliders, sleptons may
be produced (1) directly in pairs through Drell-Yan processes, (2)
singly in cascade decays of electroweak gauginos that are produced in
pairs through Drell-Yan processes, and (3) singly in squark and gluino
cascade decays.  While, for certain parameters, sleptons may be
discovered through the first two production mechanisms\cite{Baer}, the
event rates are in general too low for the precision studies considered
here.  We therefore examine the third possibility.  To do so, we begin
by considering a point in parameter space where a promising number of
sleptons are produced in gluino and squark cascades. This point,
analyzed in Refs.~\cite{SnowmassLHC,Hinchliffe}, is given by minimal
supergravity boundary conditions\footnote{Of course, strictly in minimal
supergravity, all the sleptons are exactly degenerate at $M_{\text{Pl}}$
and lepton flavor is conserved. However, given that all the sleptons
must be nearly degenerate to produce a $CP$-violating signal, we simply
use the minimal supergravity case analyzed in
Refs.~\cite{SnowmassLHC,Hinchliffe} to obtain the relevant superpartner
mass spectrum and branching ratios, which are unaffected by the small
slepton non-degeneracies and mixing angles required for our
$CP$-violating signal.}

\begin{equation}
m_0 = 100 \text{ GeV},\ m_{1/2} = 300 \text{ GeV},\
A_0 = 300 \text{ GeV},\ \tan\beta = 2.1,\ \mu>0 \ ,
\end{equation}
with resulting weak scale SUSY parameters

\begin{equation}
M_1 = 126 \text{ GeV},\ M_2 = 252 \text{ GeV},\
M_3 = 752 \text{ GeV},\ \mu = 479 \text{ GeV}\ .
\end{equation}
The more relevant masses, cross sections, and branching fractions are
given in Tables~\ref{table:masses}, \ref{table:crosssections}, and
\ref{table:branchingfractions}. At this parameter point, a large number 
of sleptons are produced through the cascades $(\tilde{g} \to )\,
\tilde{q}_L \to \neutralinotwo \to \slepton_R \to \neutralinoone$.  
The large rate results from the fact that in minimal supergravity,
$\neutralinotwo \approx \tilde{W}^3$, and $\neutralinoone \approx
\tilde{B}$, and so decays $\tilde{q}_L \to \neutralinoone$, though
favored by phase space over $\tilde{q}_L \to \neutralinotwo$, are highly
suppressed by hypercharge couplings.  Note that left-handed sleptons are
heavier than $\neutralinotwo$ and so are almost never produced.

Cuts to isolate such slepton events are presented in
Ref.~\cite{Hinchliffe}.  As noted below Eq.~(\ref{delta}), the
$CP$-violating cross section differences $\Delta_{\alpha\beta}$ in the
$e\mu$, $\mu\tau$, and $\tau e$ channels are all identical. Here we will
consider the subset of slepton events with $e^{\pm}\mu^{\mp}$ final
states and try to measure the asymmetry in this event sample.  Other
asymmetries, for example, in $e^{\pm}\pi^{\mp}$ events, where the pion
results from $\tau$ decay, also could be used.  These asymmetries suffer
from $\tau$ branching fractions, and require that hadronic jets not be
misidentified as $\tau$ decay products.  They are, however, free of
backgrounds from $\selectron$ and $\smuon$ production, whereas the
$e^{\pm}\mu^{\mp}$ asymmetry suffers from $\tilde{\tau} \to \tau\to l$
backgrounds, as discussed below.  Of course, as all of these cross
section differences $\Delta_{\alpha\beta}$ are predicted to be equal,
they could ultimately be combined to yield the most powerful measurement
of slepton $CP$ violation.

To quantify the statistical significance, we divide the signal and
background into two parts and define

\begin{equation}
A \equiv \frac{S}{B} = \frac{S^{\tilde{l}} + S^0}{B^{\tilde{l}} + B^0} \ ,
\label{asymmetrytwo}
\end{equation}
where

\begin{eqnarray}
S^{\tilde{l}} = N_{e^+_{\alpha} e^-_{\beta}}^{\tilde{l}} -
N_{e^+_{\beta} e^-_{\alpha}}^{\tilde{l}}\ , &\qquad&
S^0 = N_{e^+_{\alpha} e^-_{\beta}}^0-N_{e^+_{\beta} e^-_{\alpha}}^0\ ,\\
B^{\tilde{l}} = N_{e^+_{\alpha} e^-_{\beta}}^{\tilde{l}} +
N_{e^+_{\beta} e^-_{\alpha}}^{\tilde{l}}\ , &\qquad&
B^0 = N_{e^+_{\alpha} e^-_{\beta}}^0+N_{e^+_{\beta} e^-_{\alpha}}^0 \ . 
\end{eqnarray}
The superscripts ``${\tilde{l}}$'' and ``$0$'' denote events arising
from slepton production and events from other sources, respectively.
The statistical uncertainty in $A$ is given by $\sigma_A^2 =
(1-{A^0}^2)/B$, where $A^0 = S^0/B$ is the asymmetry in the absence of
slepton $CP$ violation.  If there is no asymmetry from other sources,
$A^0 = 0$, and an $N\sigma$ signal requires simply $|S|/\sqrt{B} > N$.

For hadron colliders, ignoring possible differences in $l^+$ and $l^-$
detection efficiencies, the source of the $CP$-violating numerator of
Eq.~(\ref{asymmetrytwo}) is entirely slepton events, and so
$S^0=0$. Since we are looking at $e^{\pm}\mu^{\mp}$ events, these
slepton events include events involving leptonically-decaying $\tau$
leptons.  The total signal is therefore

\begin{equation}
S = S^{\tilde{l}} = (S_{12} - S_{21}) \sigma_0 \varepsilon_{ll} L 
+ B_{\tau}(S_{13}-S_{31}-S_{23}+S_{32})\sigma_0 \varepsilon_{l\tau} L \ ,
\end{equation}
where $l=e, \mu$, $B_{\tau} \equiv B(\tau \to l\nu\bar{\nu}) \approx
18\%$, and $L$ is the integrated luminosity.  The efficiencies for
accepting $ll$ and $l\tau$ slepton events are denoted by
$\varepsilon_{ll}$ and $\varepsilon_{l\tau}$, and these include all
branching fractions, kinematic cut efficiencies, and detector
acceptances. Note that the contributions from $\tau$ decays
always reduce the signal $S$, since $S_{13} - S_{31} - S_{23} + S_{32} =
-\Delta_{31} - \Delta_{23} = -2\Delta_{12} = -2 (S_{12} - S_{21})$;
these processes are therefore more dangerous than other backgrounds,
which only dilute the asymmetry by increasing the background $B$.

The backgrounds include slepton events, as well as additional processes,
and are given by

\begin{eqnarray}
B^{\tilde{l}} = (\sigma_{ll} + \sigma_{l\tau} + 
\sigma_{\tau\tau}) L \\
B^0 = (\sigma_{\text{SUSY}} + \sigma_{\text{SM}}) L \ ,
\end{eqnarray}
where

\begin{eqnarray}
\sigma_{ll} &=& (S_{12} + S_{21}) \sigma_0 \varepsilon_{ll} \\
\sigma_{l\tau} &=&  
B_{\tau} (S_{13}+S_{31}+S_{23}+S_{32}) \sigma_0\varepsilon_{l\tau} \\
\sigma_{\tau\tau} &=& 2B_{\tau}^2 S_{33} \sigma_0 
\varepsilon_{\tau\tau} \\
\sigma_{\text{SUSY}} &=& \sigma_{\charginoone\charginoone} 
\varepsilon_{\charginoone\charginoone} \\
\sigma_{\text{SM}} &=& \sigma_{tt} \varepsilon_{tt} +
\sigma_{WW} \varepsilon_{WW} \ .
\end{eqnarray}
For each process, $\sigma\varepsilon$ denotes the cross section into
opposite sign, unlike flavor dilepton events, again including all
branching ratios, cuts, and detector efficiencies.  The cross section
$\sigma_{ll}$ is the irreducible background resulting directly from
slepton flavor violation, and $\sigma_{l\tau}$ ($\sigma_{\tau\tau}$) is
the cross section of $e^{\pm}\mu^{\mp}$ events resulting from slepton
production with one (two) leptonically-decaying $\tau$ lepton(s).  Other
than slepton events, the leading SUSY background process is events in
which both gluino/squark cascade decays result in charginos, which then
both decay leptonically.  These cascades are of the form $(\tilde{g} \to
) \, \tilde{q}_L \to \charginoone \to l$, and the cross section is
denoted by $\sigma_{\text{SUSY}}$ above. The leading SM backgrounds are
$t\bar{t}$ and $WW$ production, which together form
$\sigma_{\text{SM}}$.

The number of opposite sign dilepton events passing all cuts has been
plotted in Ref.~\cite{Hinchliffe}.  From these results, one may
determine the cross sections from the various sources after all cuts,
being careful to remember that opposite sign, {\em like} flavor dilepton
events resulting from $\charginoone\charginoone$, $t\bar{t}$, and $WW$
events are included in the plot, but are not to be included in our
background. For the sample of events in the range $0 < M_{ll} < 110
\text{ GeV}= M^{\text{max}}_{ll}$, where $M^{\text{max}}_{ll} = 
m_{\neutralinotwo} - m_{\neutralinoone}$ is the maximal value allowed in
slepton events, $\sigma_0\varepsilon_{ll}, \sigma_0\varepsilon_{l\tau},
\sigma_0\varepsilon_{\tau\tau} \approx 80$ fb,
$\sigma_{\charginoone\charginoone}
\varepsilon_{\charginoone\charginoone} \approx 6$ fb, and 
$\sigma_{tt} \varepsilon_{tt} + \sigma_{WW} \varepsilon_{WW} \approx 2$
fb. Note that, although the chargino background depends on a number of
additional SUSY parameters, its size may be determined by considering
{\em same} sign dilepton events. In addition, since dileptons in all of
the backgrounds have dilepton masses $M_{ll}$ that may extend beyond
$m_{\neutralinotwo} - m_{\neutralinoone}$, the size of the backgrounds
may be estimated by extrapolating from this region.

\begin{figure}
\centerline{\psfig{file=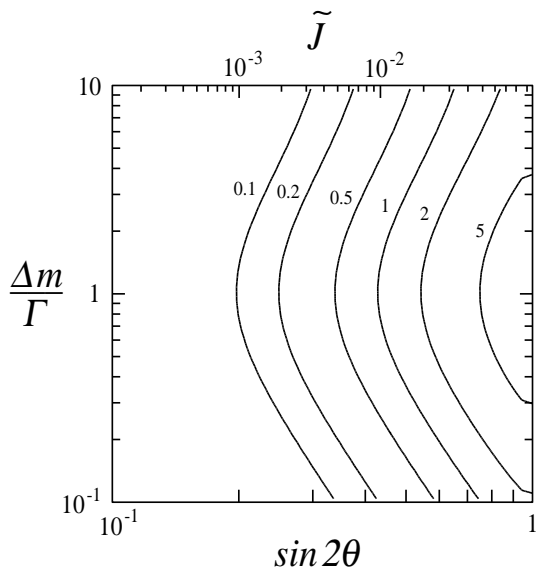,width=0.48\textwidth}}
\vspace*{.2in}
\caption{Constant contours (in fb) of the $CP$-violating cross section
$S/L$ for Scenario I at the LHC.  The signal is maximal for $\Delta m
\sim \Gamma = 0.13\text{ GeV} = 8.1 \times 10^{-4} m$. The signal $S$ is
directly proportional to $\widetilde{J}$.  For the $\sin 2\theta$ axis
the $CP$-violating phase is fixed to $\sin\delta = 1$.
\label{fig:sigmaLHC}}
\end{figure}

\begin{figure}
\centerline{\psfig{file=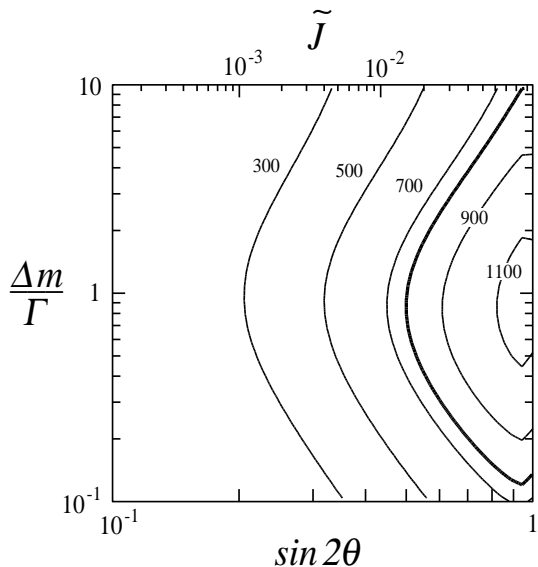,width=0.48\textwidth}}
\vspace*{.2in}
\caption{3$\sigma$ slepton $CP$ violation discovery contours at the LHC
with an integrated luminosity of 100 fb$^{-1}$ in Scenario I (two
generation degeneracy).  The $CP$-violating phase is fixed to
$\sin\delta = 1$.  The wide contour is for the parameters discussed in
the text, where $m_{\tilde{g}}= 767$ GeV, and $\Gamma = 0.13\text{ GeV}
= 8.1 \times 10^{-4} m$. The other contours give an indication of
possible discovery reaches for scenarios with $m_{\tilde{g}} =
m_{\tilde{q}}$ shown (in GeV), under the naive assumption that all cross
sections scale simply as the gluino and squark cross sections.  See full
discussion in the text.
\label{fig:LHCI}}
\end{figure}

Given these results, for each of the scenarios specified above, the
signal and background are completely determined by the parameters $x =
\Delta m/\Gamma$, $\theta$, and $\delta$. In Fig.~\ref{fig:sigmaLHC}, we
plot contours of constant $CP$-violating cross section $S/L$ for
Scenario I in the $(\sin 2\theta, \Delta m/\Gamma )$ plane for fixed
$\sin\delta=1$.  For the point we are considering, $\Gamma (\tilde{l}_R
\to l \neutralinoone) = 0.13\text{ GeV} = 8.1 \times 10^{-4} m$. We see 
that there may be a large $CP$-violating signal for $\Delta m \sim {\cal
O}(\Gamma)$.  In Fig.~\ref{fig:LHCI}, we incorporate the effects of
background and plot 3$\sigma$ discovery contours in the same plane with
$\sin\delta=1$.  The wide contour is the discovery limit for the point
we have considered, given an integrated luminosity of $100\text{
fb}^{-1}$, the luminosity expected for one detector in one year at high
luminosity. (The other contours will be described below.) The behavior
of the contours very near $\sin 2\theta = 1$ is an artifact of our
parametrization; as noted in Sec.~\ref{sec:scenarios}, $\widetilde{J}$
reaches its maximum at $\sin 2 \theta \approx 0.99$.  The overall shape
of the contour may be very roughly understood by neglecting variations
in the background and considering the behavior of the signal $S$.  For
$\Delta m \gg \Gamma$ ($x_{12} \gg 1$),

\begin{equation}
S \sim \Delta_{12} 
= 4 \sum_{i<j} \widetilde{J}_{12}^{ij} \frac{x_{ij}}{1+x^2_{ij}} 
\approx 4 \widetilde{J} \frac{x_{12}}{1+x^2_{12}}
\approx 4 \widetilde{J}/x_{12} \sim \theta^3/ \Delta m \ ,
\end{equation}
whereas for $\Delta m \ll \Gamma$ ($x_{12} \ll 1$),

\begin{equation}
S \sim \Delta_{12} \approx 4 \widetilde{J} \frac{x_{12}}{1+x^2_{12}}
\approx 4 \widetilde{J} x_{12} \sim \theta^3 \Delta m \ .
\end{equation}
For $\Delta m \approx \Gamma$, the reach in $\sin 2\theta$ is maximal,
and $\sin\theta = 0.23$ and a supersymmetric Jarlskog invariant
$\widetilde{J}$ as low as 0.01 may be probed.  We note that, while $CP$
violation requires general three generation mixing, large $CP$-violating
effects do not require all three generations to be degenerate to within
$\Gamma$.

If such a scenario were actually realized in nature, these results could
be improved with optimized cuts. For example, for slepton events,
$M_{ll}$ peaks near its maximum, and so by considering only events with
$80 \text{ GeV} < M_{ll} < 110$ GeV, the signal to background ratio is
improved.  Such a cut is also effective in removing the $\tau \to l$
backgrounds, and would lead to 
improvements in the results.

Far greater variations in these results arise if different SUSY
parameters are considered.  An obvious dependence is on the gluino and
squark masses.  For the example considered, $m_{\tilde{g}} = 767$ GeV,
and $m_{\tilde{q}_L} = 662$ GeV.  For lower masses, the cross sections
for squark and gluino production increase rapidly, and, if the relevant
branching fractions and cut efficiencies are not greatly altered, the
sample of slepton events increases dramatically.  Conversely, as the
gluino and squark masses increase, the results deteriorate.  To give an
indication of the dependence of our results on these masses, we have
also plotted 3$\sigma$ discovery contours in Fig.~\ref{fig:LHCI} for
scenarios with values of $m_{\tilde{g}} = m_{\tilde{q}}$ as labeled.
For these contours, we make the naive assumption that the signal and
background cross sections scale with $\sigma(\tilde{g}\tilde{g}) +
\sigma(\tilde{q}\tilde{q}) + \sigma(\tilde{g}\tilde{q})$ relative to 
our prototype point.  With this assumption, we see that in the favorable
case of a light gluino with mass 300 GeV, a supersymmetric Jarlskog
invariant as low as $\widetilde{J} = 10^{-3}$ may be probed.  Of course,
branching fractions and efficiencies are also highly dependent on the
various SUSY parameters, as will be discussed below, and in fact one
generally expects the difficulty of isolating SUSY signals above SM
background to increase as the superpartner masses decrease. These
contours do, however, give an indication of the strong dependence on
gluino and squark masses, and show that far stronger probes of
$\widetilde{J}$ may be possible if these masses are significantly lower.

\begin{figure}
\centerline{\psfig{file=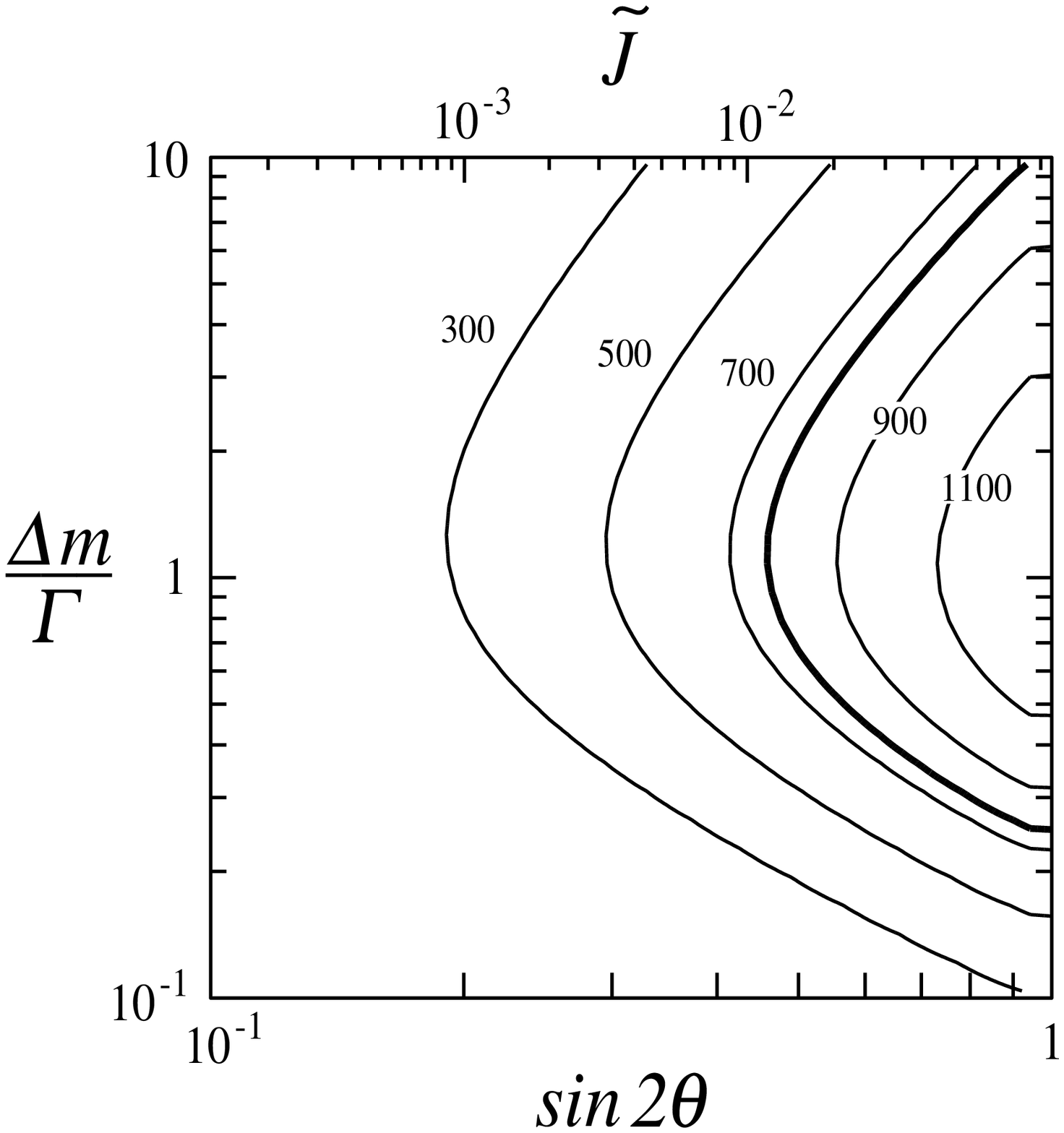,width=0.48\textwidth}}
\vspace*{.2in}
\caption{Same as for Fig.~\protect\ref{fig:LHCI}, but for Scenario II 
(three generation degeneracy).
\label{fig:LHCII}}
\end{figure}

In Fig.~\ref{fig:LHCII}, we have plotted the analogous contours for
Scenario II, the case of three generation near degeneracy.  The reach in
$\sin 2\theta$ is virtually unchanged, as is the behavior for $\Delta m
\gg \Gamma$.  For $\Delta m \ll \Gamma$, 

\begin{equation}
S \sim \Delta_{12} = 
4 \sum_{i<j} \widetilde{J}_{12}^{ij} \frac{x_{ij}}{1+x^2_{ij}} 
= 4 \widetilde{J} \left[\frac{x_{12}}{1+x^2_{12}} +
\frac{x_{23}}{1+x^2_{23}} + \frac{x_{31}}{1+x^2_{31}} \right] 
\sim \theta^3 x_{12}^3 \sim \theta^3 (\Delta m)^3 \ ,
\end{equation}
where the term linear in $\Delta m$ vanishes since $x_{12} + x_{23} +
x_{31}=0$.  For Scenario II, therefore, the signal is more strongly
suppressed for small $\Delta m$ than in Scenario I.

We have seen that the possibility of probing small $CP$-violating
parameters at the LHC exists.  However, as noted above, the size of the
$CP$-violating signal is governed by a number of factors, which we now
discuss in greater detail. First, as is obvious from the figures, given
a sample of slepton events, $CP$ violation is only significant if
$\Delta m$ is within an order of magnitude or so of $\Gamma$.  Classes
of models in which this arises naturally will be discussed below in
Sec.~\ref{sec:models}, and we therefore defer comments on the likelihood
of this possibility.

In addition, the size of the slepton event sample is governed by several
conditions on the sparticle masses and decay patterns.  For this
analysis, the strength of the LHC lies in the fact that leptonic events,
which stand out above backgrounds, are produced with strong interaction
rates.  The power of this analysis therefore depends crucially on the
number of on-shell sleptons produced in the decays of gluinos and
squarks.  We may discuss each step in the cascade $(\tilde{g} \to
)\, \tilde{q}_L \to \neutralinotwo \to \tilde{l}$.  To initiate this
decay chain, it is of course important that there be large cross
sections for gluinos and squarks; the strong dependence on their masses
was discussed above.  If we are in the gaugino region, where the
lightest two neutralinos are gaugino-like, slepton events arise from
$\tilde{q}_L$, as decays $(\tilde{g} \to )\, \tilde{q}_R \to
\neutralinoone$ bypass sleptons.  This event rate could be
much larger if, for example, $m_{\tilde{q}_R} > m_{\tilde{g}} >
m_{\tilde{q}_L}$, as then the branching fraction of $\tilde{g} \to
\tilde{q}_L$ would be nearly 100\%. In the gaugino region, it is also 
essential that $m_{\tilde{l}} < m_{\neutralinotwo}$.\footnote{If the
parameters did not lie in the gaugino region, that is, if $|\mu|$ were
not much greater than $M_{1,2}$, the heavier two neutralinos would also
have significant gaugino components, making decays $\tilde{g},\tilde{q}
\to \tilde{\chi}^0_{3,4} \to \tilde{l}$ possible, and this requirement 
would not be necessary.}  Possible decay modes of $\neutralinotwo$ are
$\neutralinotwo \to h\LSP, Z\LSP, \tilde{l}_R l,\tilde{l}_L l$.
Neglecting obvious phase space considerations, if the decay
$\neutralinotwo \to \tilde{l}_L$ is open, it will dominate, as all other
decay modes are suppressed by mixing angles in the gaugino region.  The
slepton $CP$ violation would then measure $\widetilde{J}_L$ with very
little contamination from $\widetilde{J}_R$, and with large statistics
from $B(\neutralinotwo \to \tilde{l}_L) \approx 100 \%$.  If the
$\tilde{l}_L$ decay mode is closed, but the $\tilde{l}_R$ mode is open,
we may still measure $\widetilde{J}_R$. However, this measurement is
greatly degraded if the decay $\neutralinotwo\to h$ is open, as this is
less suppressed by mixing angles in the gaugino region.

\begin{figure}
\centerline{\psfig{file=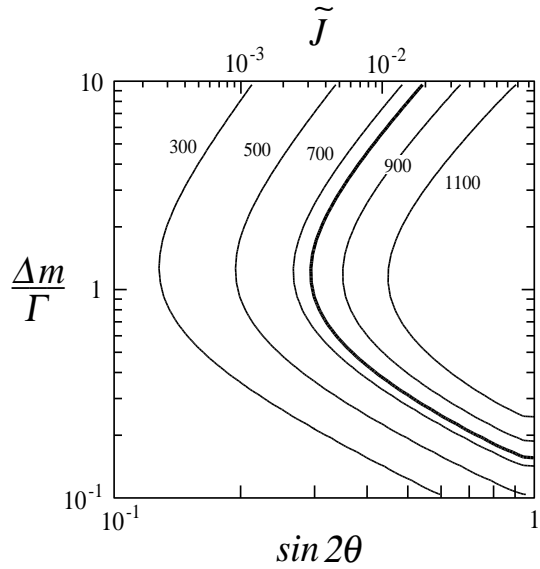,width=0.48\textwidth}}
\vspace*{.2in}
\caption{Same as for Fig.~\protect\ref{fig:LHCII}, but for 10 times
improved statistics.  Possible sources of such an improvement include
more favorable branching ratios, the combination of all lepton
asymmetries, and an increase in the assumed integrated luminosity to
include multi-year event samples and both detectors.
\label{fig:optimistic}}
\end{figure}

In the minimal supergravity example considered above, the total cross
section for gluino and squark pair production was 15.3 pb, and the
slepton cross section in the final event sample was 240 fb. Thus,
including branching ratio and kinematic cut efficiencies, only 1.6\% of
the total strongly interacting sparticle production was available for
use in our analysis.  In that example, $\neutralinotwo \to \tilde{l}_L$
decays were closed and decays $\neutralinotwo \to h$ were open, and so
$B(\neutralinotwo \to \tilde{l}_R)$ was only 36\%.  In a more favorable
scenario in which either $m_{\tilde{l}_L}<m_{\neutralinotwo}$ or
$m_{\tilde{l}_R} < m_{\neutralinotwo} < m_h$, the branching ratio to
sleptons could be nearly 3 times larger.  Clearly the results of the
previous analysis would be noticeably more powerful 
in such a scenario.  For
comparison, we present in Fig.~\ref{fig:optimistic} the results of a
scenario in which the event sample is increased relative to
Fig.~\ref{fig:LHCII} by a factor of 10.  Such an improvement could come
from improved branching ratios, the optimized cuts discussed above,
the combination of all lepton asymmetries, or by increasing the assumed
integrated luminosity to include multi-year event samples and both
detectors.

Of course, the scenario we considered was already optimistic in the
sense that sleptons were in fact produced in gluino and squark cascades.
For $m_{\tilde{l}} < m_{\neutralinotwo}$, it is likely that some probe
of slepton $CP$ violation will be possible, although the sensitivity
depends on the many issues discussed above.  However, in the gaugino
region, if $m_{\tilde{l}} > m_{\neutralinotwo}$, such an analysis will
be extremely difficult.  In this case, there is likely to be little
opportunity for high precision studies of sleptons at the LHC. Sleptons
may be produced in large numbers at the NLC, however, and this
production is direct, that is, independent of various cascades and
branching ratios.  We therefore now turn to a scenario in which slepton
$CP$ violation is undetectable at the LHC and consider the sensitivity
of similar probes at the NLC.

\subsection{NLC}

At $e^+e^-$ colliders, sleptons are dominantly produced directly through
correlated pair production.  Relative to the situation at hadron
colliders, slepton production therefore is not as heavily influenced by
details of the sparticle spectrum and branching ratios.  We will see
that, if slepton pairs are kinematically accessible, $e^+e^-$ colliders
offer a robust opportunity to measure slepton $CP$ violation.  Sleptons
pairs $\tilde{l}_L\tilde{l}_L$ and $\tilde{l}_R\tilde{l}_R$ may be
produced through $s$-channel photon and $Z$ diagrams, and also through
$t$-channel neutralino exchange.  $CP$ asymmetries in
$\tilde{l}_L\tilde{l}_L$ and $\tilde{l}_R\tilde{l}_R$ events measure
$\widetilde{J}_L$ and $\widetilde{J}_R$,
respectively.\footnote{Associated production of $\tilde{l}_L\tilde{l}_R$
through $t$-channel neutralino exchange is also possible.  This is
uncorrelated production of sleptons at separate vertices, and so the
$CP$ asymmetry in this process is a linear combination of
$\widetilde{J}_L$ and $\widetilde{J}_R$.}  In this subsection, we will
study the prospects for $e^+e^-$ colliders by considering the
experimental setting of the NLC, a proposed high energy linear $e^+ e^-$
collider, with maximum center-of-mass energy $\sqrt{s} = 0.5$--1.5 TeV
and luminosity ${\cal L}
\sim 50-100$ fb$^{-1}$/yr.  At such a collider, highly polarized $e^-$
beams are expected to be available.

For the NLC, we study a scenario specified by weak-scale SUSY parameters

\begin{eqnarray}
m_{\tilde{l}_R} = 200 \text{ GeV},\ m_{\tilde{l}_L} &=& 350 \text{ GeV},\
M_2 = 2 M_1 = 190 \text{ GeV}, \nonumber \\
\tan\beta=2,\ && \mu=-400 \text{ GeV} \ .
\end{eqnarray}
In this scenario, the lightest two neutralinos are gaugino-like and the
decay $\neutralinotwo \to \tilde{l}$ is forbidden.  Thus, sleptons are
not produced in gluino and squark cascades at the LHC, and precision
studies of sleptons there are likely to be very difficult.

With the underlying SUSY parameters above and $\sqrt{s} = 500$ GeV,
right-handed sleptons are pair-produced and decay through $\tilde{l}_R
\to l \neutralinoone$. Decays $\tilde{l}_R \to \nu_l \charginoone, 
l\neutralinotwo$ are also allowed, but are highly suppressed by phase
space and mixing angles, since the chargino and neutralino are SU(2)
gaugino-like.  Note that both $\tilde{l}_L \tilde{l}_R$ and $\tilde{l}_L
\tilde{l}_L$ production are inaccessible at $\sqrt{s} = 500$ GeV.  Thus,
the effects of $CP$ violation in the $\tilde{l}_L$ sector are completely
removed, and all slepton $CP$ violation is a measure of
$\widetilde{J}_R$.  Of course, in any fortunate scenario in which both
left- and right-handed sleptons may be produced, the beam energy may
always be tuned to eliminate production of either $\tilde{l}_L$ or
$\tilde{l}_R$, with possibly significant loss in slepton cross section.
If flavor and $CP$ violation in the $\tilde{l}_R$ sector is
well-understood, one could then increase the beam energy and attempt to
measure the additional parameters that enter with the production of
left-handed sleptons.

Cuts have been studied in Ref.~\cite{BV} for slepton flavor-conserving
signals.  We adopt those cuts here for the case of slepton
flavor-violating signals.  The efficiency of these cuts for the signal
is approximately 30\%.\footnote{The 30\% efficiency is for $\tilde{e}$
and $\tilde{\mu}$ events from Ref.~\cite{BV}.  For the purposes of
calculating background, we will assume this efficiency also for events
involving $\tilde{\tau}$.}  The dominant SM backgrounds are $WW$, $e\nu
W$, and $eeWW$.  With unpolarized beams, the SM background cross section
is 4.8 fb. However, as in our previous study\cite{lfv}, we may exploit
the advantages of right-polarized beams, which, in this case, doubles
our signal cross section and reduces the SM background to 2.6 fb for
90\% beam polarization.

We must also consider the additional SUSY backgrounds.  For this beam
energy, $\neutralinoone \neutralinotwo$ and $\neutralinotwo
\neutralinotwo$ production are kinematically accessible, but these 
are suppressed to negligible levels by mixing angles in the gaugino
region.  Chargino pair production is also allowed, and may produce
dilepton events with large acoplanarity and missing $p_T$.  However, the
chargino background is reduced by the branching ratio
$B(\tilde{\chi}^{\pm}_1 \to l\nu \neutralinoone)$.  In addition, the
$e^-_R$ beam polarization strongly suppresses chargino production, as
the charginos are almost purely gaugino.  For 90\% beam polarization,
the chargino background is suppressed to a low level relative to the
leading SM backgrounds, and may be safely ignored in the following
analysis.\footnote{If left-handed sleptons are to be studied, chargino
pair background cannot be removed with beam polarization.  However, for
$m_{\charginoone} > m_{\tilde{l}_L}$, the beam energy could be tuned to
eliminate chargino pair production.}

The expressions for $S^{\tilde{l}}$ and $B^{\tilde{l}}$ are as in the
previous subsection, with the exception that the correlated pair
expressions are to be used.  For the other background processes,

\begin{eqnarray}
S^0 &=& 0.6 \text{ fb} \times L \\
B^0 &=& 2.6 \text{ fb} \times L \ . 
\end{eqnarray}
In contrast to the LHC case, $S^0$ is not zero --- the beam polarization
induces an asymmetry in the SM background, since $e^+e^-_R \to e^-\nu
W^+$ is allowed, but $e^+e^-_R \to e^+\nu W^-$ is forbidden.  It is
therefore important that the beam polarization and SM asymmetry be
well estimated.

\begin{figure}
\centerline{\psfig{file=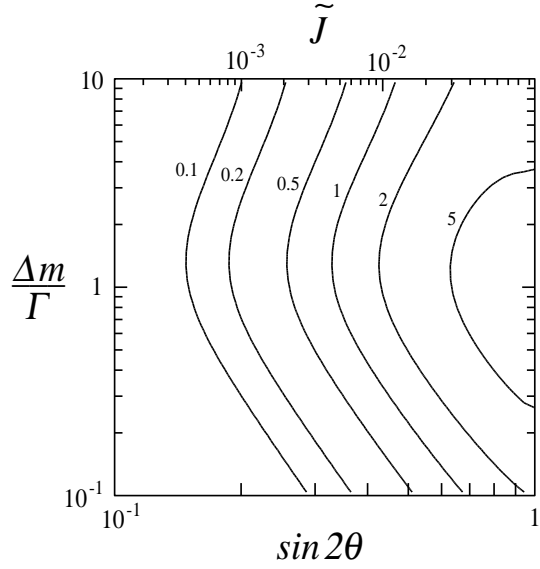,width=0.48\textwidth}}
\vspace*{.2in}
\caption{
Constant contours (in fb) of the $CP$-violating cross section $S/L$ for
Scenario I at the NLC.  The signal is maximal for $\Delta m \sim \Gamma
= 0.58\text{ GeV} = 2.9 \times 10^{-3} m$. The $CP$-violating phase is
fixed to $\sin\delta = 1$.
\label{fig:sigmaNLC}}
\end{figure}

\begin{figure}
\centerline{\psfig{file=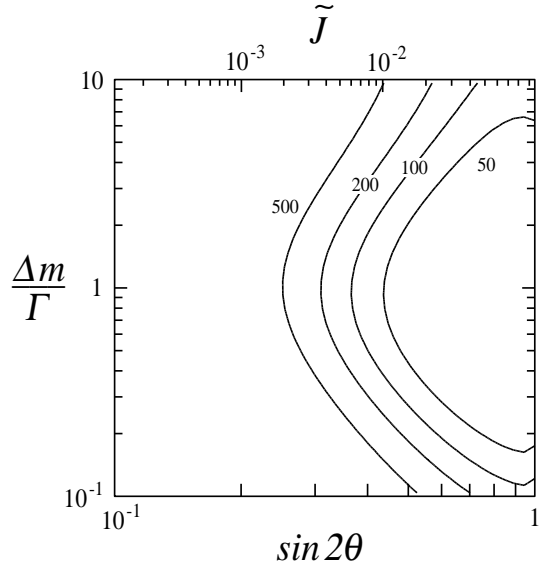,width=0.48\textwidth}}
\vspace*{.2in}
\caption{3$\sigma$ slepton $CP$ violation discovery contours at the NLC
with integrated luminosity given (in fb$^{-1}$) for Scenario I (two
generation degeneracy).  The $CP$-violating phase is fixed to
$\sin\delta = 1$.  The SUSY parameters are as given in the text, with
$\Gamma = 0.58\text{ GeV} = 2.9 \times 10^{-3} m$.
\label{fig:NLCI}}
\end{figure}

\begin{figure}
\vspace*{.2in}
\centerline{\psfig{file=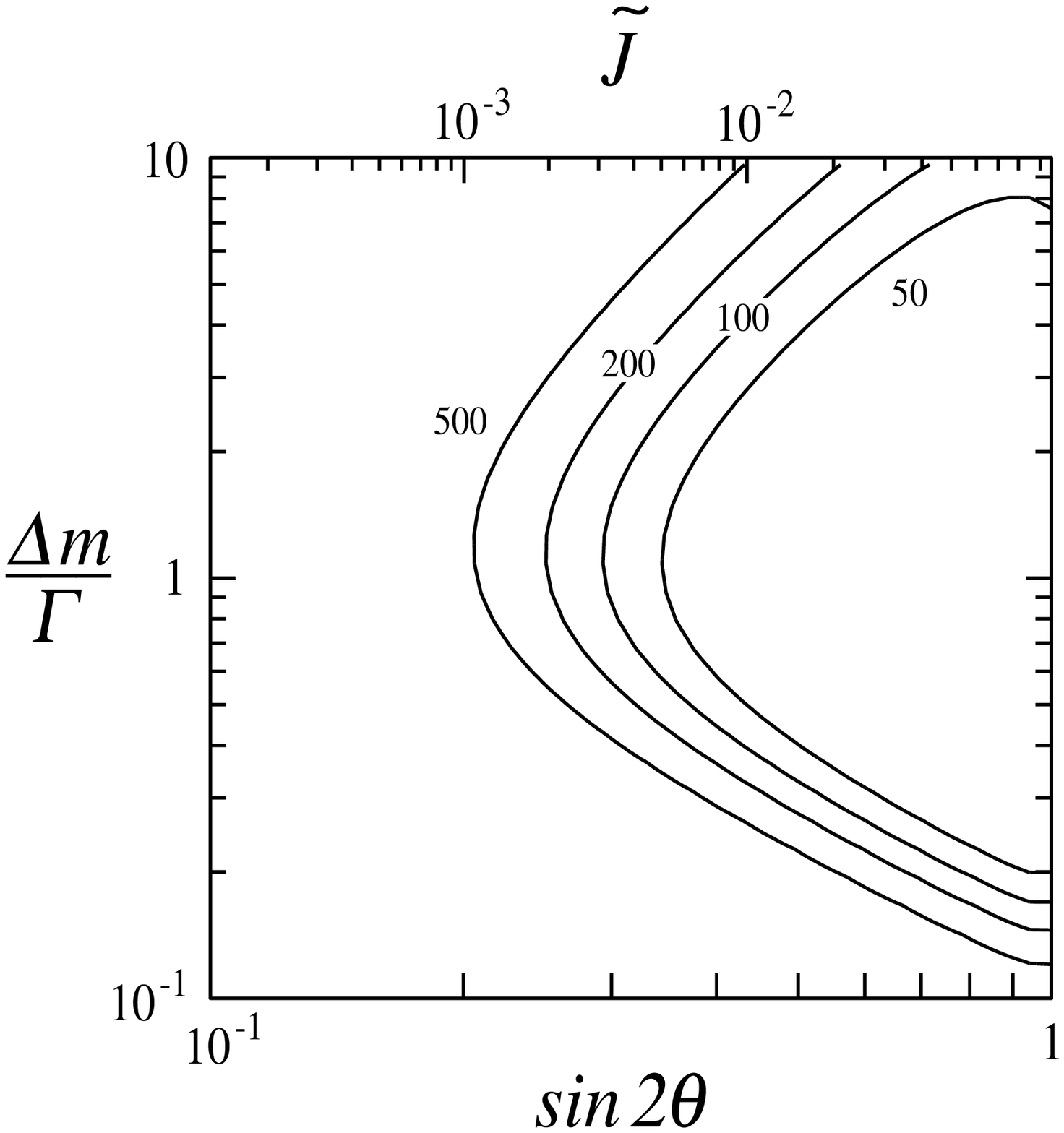,width=0.48\textwidth}}
\caption{Same as for Fig.~\protect\ref{fig:NLCI}, but for Scenario II 
(three generation degeneracy).
\label{fig:NLCII}}
\end{figure}

Contours of constant $CP$-violating cross section $S/L$ for Scenario I
are plotted in Fig.~\ref{fig:sigmaNLC} in the $(\sin 2\theta, \Delta
m/\Gamma )$ plane for fixed $\sin\delta=1$.  The $CP$ asymmetry is
maximal for $\Delta m \approx \Gamma$, and for $\widetilde{J} \approx
10^{-2}$, the difference can be as large as 100 events per year. In
Figs.~\ref{fig:NLCI} and \ref{fig:NLCII} the discovery reach at the NLC
for various integrated luminosities is plotted for Scenarios I and II.
Relative to the LHC, the contour shapes are more difficult to
understand, as the expressions for correlated pair production are
complicated, but the essential behavior is similar to that of the LHC
case.  The maximal reach in $\sin 2\theta$ is for $\Delta m \sim \Gamma
= 0.58 \text{ GeV} = 2.9 \times 10^{-3} m$, and values of $\widetilde{J}
\approx 10^{-3} - 10^{-2}$ may be probed, depending on the integrated
luminosities.

For large integrated luminosities, the NLC may therefore also probe
small $CP$-violating parameters.  Of course, for SUSY parameters ideally
suited to the LHC, where gluinos and squarks are light and decay
frequently to sleptons, the sensitivity at the NLC is not competitive
with that at the LHC.  This stems from the fact that the slepton
production mechanism at the NLC is through weak interactions, whereas
that at the LHC is through strong interactions. The strength of the NLC,
however, is that the analysis does not rely on favorable gluino and
squark masses and decay patterns.  For example, if gluino and squark
masses are very large, or sleptons are simply not produced in squark and
gluino cascade decays, slepton studies may be extremely difficult at the
LHC, but would still be possible at the NLC.  In addition, in some
instances, the NLC provides the possibility of disentangling left- and
right-handed slepton flavor and $CP$ violation by gradually raising the
beam energy.  The NLC therefore provides a robust probe of slepton $CP$
violation, for the most part requiring only that slepton pairs be
kinematically accessible.

\section{Constraints from the electron EDM}
\label{sec:edm}

We have seen in the previous sections that two conditions are required
for a large $CP$-violating signal in the oscillation of sleptons: all
the mixing angles in the $W$ matrix should be large (so that
$\widetilde{J}$ is large), and at least two of the sleptons should be
degenerate to nearly their widths. 
In this section, we discuss the contribution to the electron
EDM coming from the $CP$-violating phases in the $W$ matrices. As
already mentioned in the discussion of $CP$-violating rephase
invariants, the dominant contributions to the electron EDM due to the
phases in $W$ require non-trivial mixing in both the left and right
slepton sectors, and are thus probing the invariants
$\widetilde{K}_{12,13}$, which are completely distinct from the
invariants $\widetilde{J}_{L,R}$ probed at
colliders. Nevertheless, we wish to show that in the case when all
$CP$-violating phases contributing to the $\widetilde{J}$'s and
$\widetilde{K}$'s are comparable, and even if the bound on the electron
EDM improves by a factor of 10, a large $CP$-violating signal can still
be visible at the LHC and NLC. This is due to the fact that, for
sleptons degenerate enough for visible collider $CP$ violation, the
constraint on the $\widetilde{K}$ from $d_e$ places almost no restriction
on the size of the mixing angles, and so does not limit the size of the
collider $CP$-violating signal.

Consider the contribution to $d_e$ coming from
$\widetilde{K}_{12,13}$. For simplicity, we compute the contribution
{}from each of them separately, {\em i.e.}, we first assume only two
generation mixing and compute the contribution coming from
$\widetilde{K}_{12}$, then assume that the first two generation sleptons
are exactly degenerate and compute the contribution from
$\widetilde{K}_{13}$.  In the following we assume that the lightest
neutralino is primarily $\tilde{B}$ with mass $M_1$.  The contribution
from $\widetilde{K}_{12}$ is then

\begin{equation}
{d^{12}_e \over e} = -\widetilde{K}_{12} \times {\alpha_1 \over {4 \pi}}
{m_\mu (A - \mu \tan \beta)\over {M_1^3}} \left({\Delta
m^2_{12}\over \bar{m}_{12}^2}
\right)_L \left({\Delta m^2_{12}\over \bar{m}_{12}^2}
\right)_R f(x_L,x_R) \ ,
\end{equation}
where $\Delta m^2_{12} = m_1^2 - m_2^2$, $\bar{m}_{12}^2 = (m_1^2 +
m_2^2)/2$, $x = \bar{m}_{12}^2/M_1^2$, and

\begin{equation}
f(x_L,x_R)=x_L x_R {\partial ^2 \over{\partial x_L \partial x_R}}
\left({g(x_L) - g(x_R) \over {x_L - x_R}}\right), \; \;g(x) = {x^2 - 2 
x \ln x - 1
\over {2 (x - 1)^3}} \ .
\end{equation}
Numerically, $f(1,1) =-1/30$. The contribution from $\widetilde{K}_{13}$
is enhanced by $m_\tau/m_\mu$ relative to the one from
$\widetilde{K}_{12}$:

\begin{equation}
\label{bigeq}
{d^{3}_e \over e} = -\widetilde{K}_{13} \times {\alpha_1 \over {4 \pi}}
{m_\tau (A - \mu \tan \beta) \over {M_1^3}} \left({\Delta
m^2_{12-3}\over \bar{m}_{12-3}^2}
\right)_L \left({\Delta m^2_{12-3}\over \bar{m}_{12-3}^2}
\right)_R f(x_L,x_R) \ ,
\end{equation}
with $\Delta m^2_{12-3} = m_3^2 - m_0^2, \bar{m}_{12-3}^2 = (m_3^2 +
m_0^2)/2$, where $m_0$ is the common mass for the first two generations,
and here $x=\bar{m}^2_{12-3}/M_1^2$. (Actually, since all three
generations need to be approximately degenerate, the slepton masses in
the definition of $x$ can be taken as the average of all the scalar
masses.)

We know that in order to have a large signal from our direct probe of
the $\widetilde{J}$, we need the sleptons degenerate to within widths,
so we expect $\Delta m^2 / m^2 \sim 10^{-3} - 10^{-2}$. Requiring the
induced EDM to be less than the current experimental bound of $4\times
10^{-27} e$ cm \cite{edmref} by a factor of 10 yields the bounds

\begin{eqnarray}
\left|\widetilde{K}_{12}\right| &<& 3\times 
\left[\frac{M_1}{200 \text{ GeV}}\right]^2
\left|\frac{M_1}{A - \mu \tan \beta}\right| 
\left[\frac{f(1,1)}{f(x_L,x_R)}
\right] \left|\frac{10^{-2}}{\left(
\frac{\Delta m^2_{12}}{m^2_{12}}\right)_L}\right|
\left|\frac{10^{-2}}{\left(
\frac{\Delta m^2_{12}}{m^2_{12}}\right)_R}\right| \nonumber \\
\left|\widetilde{K}_{13}\right| &<& 0.2 \times 
\left[\frac{M_1}{200 \text{ GeV}}\right]^2
\left|\frac{M_1}{A - \mu \tan \beta}\right| 
\left[\frac{f(1,1)}{f(x_L,x_R)}
\right] \left|\frac{10^{-2}}{\left(
\frac{\Delta m^2_{12-3}}{m^2_{12-3}}\right)_L}\right|
\left|\frac{10^{-2}}{\left(
\frac{\Delta m^2_{12-3}}{m^2_{12-3}}\right)_R}\right| \ .
\end{eqnarray}

These constraints do not place significant restrictions on the sizes of
the mixing angles (unless $\tan\beta$ is known to be large) and hence do
not affect the reach for our signal from the $\widetilde{J}$.  Of
course, if an electron EDM is discovered, we cannot determine whether it
is due to the $\widetilde{K}$; it could dominantly be due to a relative
phase between the $\tilde{B}$ mass and the $A$ parameter, for
instance. Thus, a non-zero $d_e$ does not necessarily imply non-zero
$\widetilde{K}_{12,13}$. However, the $CP$ asymmetries between lepton
flavor-violating events we discussed in the previous section must be due
to $\widetilde{J}_{L,R}$, and seeing these asymmetries means that at
least one of the $\widetilde{J}_{L,R}$ is non-zero.

\section{Constraints from $\protect\bbox{\mu \to \lowercase{e} \gamma}$}
\label{sec:mutoegamma}

The large mixing angles in the $W$ matrices needed for a large
$CP$-violating signal may give dangerously large contributions to the
rate for $\mu \to e \gamma$.  We know, however, that a large
$CP$-violating signal requires at least two of the sleptons to be
degenerate to within widths. In this section, we show that as long as
the third generation sleptons also have small splittings from the other
two generations, the slepton degeneracy is sufficient to suppress the
$\mu \to e \gamma$ amplitude to harmless levels, even if there are large
mixing angles in the $W$ matrices.

There are many contributions to $\mu \to e \gamma$, but they fall into
two classes: those which require flavor mixing for both left- and
right-handed sleptons, and those which require only left- or
right-handed mixing. The most dangerous case is when there is mixing
with the third generation for both left- and right-handed scalars;
insertion of the left-right mass for the stau gives a contribution to
the $\mu \to e \gamma$ amplitude proportional to $m_\tau$, while all
other contributions are at most proportional to $m_\mu$. If we assume
for simplicity that the first two generation sleptons are exactly
degenerate and keep only the $\tilde{B}$ intermediate gaugino, then with
the same notation as in Eq.~(\ref{bigeq}) the branching ratio for $\mu
\to e \gamma$ relative to the current experimental bound of $4.9 \times
10^{-11}$\cite{muegam} is

\begin{eqnarray}
{B(\mu \to e \gamma) \over {4.9\times 10^{-11}}} &=& \left[
\frac{200 \text{ GeV}}{M_1} \right]^4 \left|{A - \mu \tan \beta \over M_1}
\right|^2 \left[\frac{f(x_L,x_R)}{f(1,1)}\right]^2\nonumber \\
&\times&\left|
\frac{\left(W_{32}W^*_{33} \frac {\Delta m^2_{12-3}}{m^2_{12-3}}
\right)_L}{5\times 10^{-2}}\right|^2
\left|
\frac{\left(W_{31}^* W_{33} \frac {\Delta m^2_{12-3}}{m^2_{12-3}}
\right)_R}{5\times 10^{-2}}\right|^2,
\end{eqnarray}
plus a similar contribution with $L \to R$.  We see that, if all the
mixing angles $W_{L,R}$ are large, we cannot allow ${\cal O}(1)$
splittings between the third generation and first two generations for
both left- and right-handed sleptons. If, however, the splittings
are less than $\sim 10^{-1}-10^{-2}$ as in our Scenario I and II, it
does not constrain the size of the mixing angles.

We now turn to contributions needing either only left-handed or only
right-handed mixing. There are again many diagrams. 
Let us consider only the contribution from the one involving the
left-right mass insertion for the $\tilde{\mu}$. Since it is
proportional to $A - \mu \tan \beta$, this contribution begins to
dominate for moderately large $\tan \beta$ and is otherwise comparable
to the other contributions. We find that

\begin{eqnarray}
{B(\mu \to e \gamma) \over {4.9\times 10^{-11}}} &=& \left[
\frac{200 \text{ GeV}}{M_1} \right]^4 \left|{A - \mu \tan \beta \over M_1}
\right|^2 \left[\frac{h(x_L,x_R)}{h(1,1)}\right]^2\nonumber \\
&\times&\left|
\frac{\left(W_{32}W^*_{31} \frac {\Delta m^2_{12-3}}{m^2_{12-3}} + 
 W_{22} W^*_{21} \frac {\Delta m^2_{12}}{m^2_{12}}\right)_L}{
2\times 10^{-2}}\right|^2.
\end{eqnarray}
with a similar contribution from $L \to R$, and where

\begin{equation}
h(x_L,x_R) = x_L \frac{\partial}{\partial x_L} 
\frac{g(x_L)-g(x_R)}{x_L - x_R} \ , 
\end{equation}
with $h(1, 1)=1/20$.

{}From the above we see that even if only pure $L$ or $R$ mixing
contributions are considered, if
all the mixing angles are large, no ${\cal O}(1)$ splittings are allowed
between any pair of sleptons. On the other hand, if (as in our Scenario
II for large $CP$ violation) all the sleptons are degenerate to within
their widths (so that $\Delta m^2/m^2 \sim 10^{-3}-10^{-2}$), the
contributions to $\mu \to e \gamma$ are already suppressed enough even
for maximal mixing angles. Even if only the first two generation 
sleptons are
degenerate within widths, and the third is split off by a factor of 10
compared to the width (as in our Scenario I), the bounds from $\mu \to e
\gamma$ can still be met with moderately small products of mixing angles
$W_{32}W_{31}^{*} \sim 1/10$, which does not place a 
strong constraint on the size of $\tilde{J}$. For large 
$W_{22}W^{*}_{21}$ (since its size is not restricted if 
$\Delta m_{12}\sim \Gamma$) and $\sin \delta$, $\tilde{J}$ can still
be as large as a few $\times 10^{-2}$, well within the reach of
the experiments.

\section{Implications for Distinguishing Models for Scalar Masses}
\label{sec:models}

{}From the discussion in the previous sections, we see that an
observable $CP$-violating signal requires the following conditions.
First, the slepton mixing angles and $CP$-violating phase should be
quite large to give a sufficiently large $\widetilde{J}$, and at least
two generations of sleptons should be degenerate to about their widths
to maximize the signal.  In addition, constraints from the flavor
changing process $\mu
\to e\gamma$ require that all three generations of sleptons should be
quite degenerate if all mixing angles are large.  These considerations
strongly suggest a particular pattern for the slepton mass matrix. Since
the sleptons must be nearly degenerate, their mass matrix must be very
nearly proportional to the identity. However, since all the mixing
angles must be large and the mass splittings must be small, the slepton
mass matrix must deviate from the identity by small perturbations to all
the entries. The mass matrix must therefore have the form

\begin{equation}
m^2_{ij} = m_0^2 \left(\delta_{ij} + \epsilon_{ij}\right) \ ,
\end{equation}
where the $\epsilon_{ij}$ are of order $\Gamma / m$, and are of
comparable size for all $i,j$.  What sort of models can give rise to
this sort of mass matrix? The degeneracy among different generations of
sleptons can originate from dynamics, flavor symmetry, or both. In the
following, we discuss how this sort of slepton mass matrix can arise in
these different cases.

First, let us consider a case where no particular flavor symmetry is
present, and the sleptons are degenerate because the mediation of SUSY
breaking to the ordinary fields is flavor-blind.  Models with SUSY
breaking mediated by gauge interactions are examples of this
class\cite{GM}. In this case, there are two important scales relevant to
our study: $\Lambda_S$, the messenger scale of SUSY breaking, where SUSY
breaking is first transmitted to the ordinary sector, and $\Lambda_D$,
the scale where extra non-universal and flavor-violating interactions
decouple. If $\Lambda_D > \Lambda_S$, the flavor-blind mediation of SUSY
breaking will guarantee the degeneracy of sleptons, and hence render the
slepton mixing matrices $W_{L,R}$ trivial. On the other hand, suppose
$\Lambda_D \alt \Lambda_S$, and the sleptons have flavor-violating
interactions $h_{iI\alpha} l_i \bar{F}_I \phi_{\alpha}$, where
$h_{iI\alpha}$ are Yukawa couplings, $l_i$ is the generation $i$
slepton, and $\bar{F}$ and $\phi$ are extra fields decoupling at
$\Lambda_D$.  The slepton masses then receive extra non-universal,
flavor-violating corrections
\begin{equation}
\Delta m^2_{ij} = 
-\frac{1}{8 \pi^2} \sum_{I \alpha} 
h_{i I \alpha} h^*_{j I \alpha}
\left(m_0^2 + m_{\phi_\alpha}^2 + m_{\bar{F}_I}^2\right) 
\left(\ln \frac{\Lambda_S}{\Lambda_D}+c_{\text{TH}}\right) \ ,
\label{correction}
\end{equation}
where $c_{\rm TH}$ represents the threshold corrections. If the
couplings $h_{iI\alpha}$ are complex and $\cal O$(1) (since there is no
flavor symmetry to restrict their sizes), the mixing matrix that
diagonalizes the slepton mass matrix will have large mixing angles and a
$\cal O$(1) $CP$-violating phase. If the logarithm is not large, then
the splittings among different generations are small, naturally of the
order $\Delta m /m \sim h^2/16 \pi^2 \sim 10^{-2}-10^{-4}$ for $h \sim
0.1-1$, which is comparable to the slepton width $ \Gamma /m \sim
g_1^2/8 \pi \times \text{phase space factor}$, as required for an
observable $CP$-violating signal. The closeness of $\Lambda_D$ and
$\Lambda_S$ could be accidental, or may arise because they have a common
origin.\footnote{For instance, they may both be triggered by the SUSY
breaking. Such a framework is discussed in Ref.~\cite{ACHM} within the
context of flavor symmetries.} Another possibility is that the dominant
contribution to the scalar masses is flavor-blind, say, from gauge
mediated SUSY breaking, but the fundamental SUSY breaking vacuum energy
$F^2$ is large enough ($F \sim 10^{18}-10^{20} \text{ GeV}^2$) that the
supergravity contribution to the scalar mass squared $\sim
(F/M_{\text{Pl}})^2 \sim (1-10 \text{ GeV})^2$ is about 
$10^{-4}-10^{-2}$ of the gauge-mediated contribution. If the
supergravity mediation does not obey any flavor symmetry, we expect
comparable contributions to all elements of the scalar mass matrices,
again leading to the small non-degeneracies and large mixing angles
needed for the $CP$-violating signal.

In the case when there is a flavor symmetry that is broken at a scale
$\Lambda_F$ and gives rise to the observed pattern of fermion masses,
the degeneracy among different generation sleptons can either come from
the dynamics or be ensured by the flavor symmetry itself (for certain
non-Abelian flavor symmetries).  Beneath the flavor symmetry breaking
scale $\Lambda_F$, extra flavor-violating interactions can be present
down to $\Lambda_D$ if not all the fields decouple at $\Lambda_F$.  We
may still get the radiative corrections to the slepton masses given in
Eq.~(\ref{correction}) for $\Lambda_D \alt \Lambda_S$, but some of the
couplings $h_{iI\alpha}$ will be small because of the flavor symmetry.
In particular, we do not expect that both $l_i$ and $l_j$, for $i\neq
j$, can have $\cal O$(1) couplings to the same fields $\bar{F}, \phi$ if
$l_i$ and $l_j$ transform differently under the flavor symmetry. The
off-diagonal elements of the mass matrix generated from radiative
corrections will then be suppressed.  The splitting generated in the
diagonal elements may be less suppressed or unsuppressed because the
different generations may have different couplings, if, for example, the
flavor symmetry is Abelian.  However, if $\Lambda_F < \Lambda_S$, there
will be additional corrections to the scalar masses from the flavor
symmetry breaking effects. These corrections come from flavor symmetric
operators like

\begin{equation}
m_S^2 {{\phi^i}^{\dagger} \over M} l_i^{\dagger} {{\phi^j} \over M} l_j\ ,
\label{operator}
\end{equation}
where the $\phi$ fields are SM singlets that carry flavor (``flavons"),
and with which the sleptons combine to form flavor singlets.  These
non-renormalizable operators may be present in the fundamental theory as
$M_{\text{Pl}}$ suppressed operators (in which case $M=M_{\text{Pl}}$),
or can arise in the Froggatt-Nielsen mechanism\cite{FN} when the heavy
Froggatt-Nielsen fields are integrated out (in which case
$M=M_{\text{FN}}$). When the flavon fields $\phi$ acquire nonzero vacuum
expectation values, breaking the flavor symmetry, the operators of
Eq.~(\ref{operator}) become mass terms for the sleptons. They can
contribute to both the off-diagonal elements and the splitting in the
diagonal elements, depending on the details of particular models. The
(spontaneously broken) flavor symmetry is expected to explain the light
fermion masses: small Yukawa couplings are understood as powers in the
small parameters $\langle \phi \rangle / M$.  In many theories, the
parameters $\langle \phi \rangle / M$ are of the order $10^{-1}-10^{-2}$
in order to generate the observed fermion mass hierarchy. The slepton
mass splitting induced by Eq.~(\ref{operator}) could be close to their
widths for suitable powers of $\phi$ in these operators. In addition,
large $CP$-violating phases in the slepton mixing matrices are expected
to be generated in this case since similar operators also give rise to
the $CP$-violating phase in the CKM matrix.  Thus, in theories in which
spontaneously broken flavor symmetries are used to explain the fermion
mass hierarchy, the necessary conditions for generating a large
$CP$-violating signal may also arise, as long as the scale of symmetry
breaking is beneath the scale at which SUSY breaking is transmitted to
the sfermions.

\section{Conclusions}
\label{sec:conclusions}

In this paper, we have studied the possibility of probing $CP$ violation
in the supersymmetric lepton mixing matrices $W_{L,R}$ which diagonalize
the left- and right-handed slepton mass matrices in the basis where the
charged lepton masses are diagonal.  There are four independent
$CP$-violating phases associated with the slepton mass matrices, and we
gave them a rephase invariant description in terms of the four quantities
$\widetilde{J}_{L,R}$ and $\widetilde{K}_{12,13}$. While the electron
EDM depends on the $\widetilde{K}$, the $CP$-violating asymmetries in
the oscillations and decays of sleptons at colliders depend on
$\widetilde{J}$, and are not directly constrained from the bounds on the
electron EDM. The $CP$-violating effects associated with a single
$\widetilde{J}$ disappear both as $\widetilde{J}$ goes to zero as well
as when any two slepton masses become degenerate. We found that three
conditions are necessary to produce a large collider $CP$ asymmetry in
lepton flavor-violating events:
\begin{enumerate}
\item The $W$ matrices must have large $CP$ and flavor-violating angles,
leading to $\widetilde{J} > 10^{-3}$.
\item The two sleptons of highest degeneracy should have $\Delta m$
within an order of magnitude of $\Gamma$.
\item All three sleptons must have some degree of degeneracy,
{\em i.e.}, $\Delta m/m \ll 1$.
\end{enumerate}
The last requirement is necessary to ensure that the large flavor mixing
angles of $W$ do not violate bounds from $\mu\to e \gamma$. It is
interesting that the observation of a sizeable $CP$-violating signal
tells us so much about the structure of the slepton mass matrix in a
region where the sleptons are so degenerate that the limits from rare
processes and EDMs are irrelevant. This is because we are probing an
oscillation phenomenon, and the superGIM suppression of the amplitude
due to near slepton degeneracy occurs only for $\Delta m/\Gamma \ll 1$,
rather than $\Delta m/m \ll 1$ as in the low energy rare processes. 

We showed that the experimental probe of $\widetilde{J}$ at both the LHC
and NLC can be powerful. Our signal at the LHC is from strong production
of squarks and gluinos, followed by cascade decays to the second
neutralino, which in turn decays into slepton and lepton, with the
slepton subsequently oscillating and decaying into the lightest
neutralino and lepton. If such cascade decays are realized in nature, we
find that the LHC can probe significant regions in the $(\widetilde{J},
\Delta m/\Gamma)$ plane. The reach depends sensitively on the strong
production cross section and hence on the squark and gluino masses, but
we find that $\widetilde{J} > 10^{-3}$ and $\Delta m/\Gamma$ between
1/10 and 10 can be probed at the LHC.  If slepton $CP$ violation is
undetectable at the LHC, a robust probe is provided by the NLC, where
sleptons are produced in great numbers independent of the various
cascades and branching ratios required at the LHC. We find that the NLC
has a comparable reach in the $(\widetilde{J}, \Delta m/\Gamma)$ plane
for high integrated luminosities.  In this paper, we have concentrated
on signals at the LHC and NLC, because these machines offer the
possibility of probing $\widetilde{J}$ to the level of $10^{-3}$ or
perhaps even further, well below the maximum of value of $\frac{1}{6
\sqrt{3}}$. We also noted, however, that if slepton masses are near
their current bound and $\widetilde{J}$ is near maximal, slepton $CP$
violation may also be observable at LEP II.  Similarly, if squarks and
gluinos are very light, a possibility of detecting slepton $CP$
violation may exist at Run II and future upgrades of the Tevatron.

We finally considered the implications of an observable $CP$-violating
signal for models of the scalar masses.  If a $CP$-violating asymmetry
is observed, we saw that the mass splittings must be non-zero and $\sim
\Gamma$.  It should be noted that the sensitivity of the $CP$-violating
signal to small but non-zero mass splittings is extraordinary. In the
examples we gave, mass splittings as small as $\sim 50$ MeV could be
distinguished from zero; such precision for slepton mass differences is
typically unobtainable through conventional considerations of kinematic
distributions.  We saw that an observable $CP$-violating signal implies
further that the scalar masses are dominantly universal, but have extra,
non-hierarchical contributions to all elements of the mass matrix that
are $\sim \Gamma/m \sim 10^{-2}-10^{-4}$ of the universal mass.  Such a
structure for scalar mass matrices does not typically arise in the
simplest models for solving the supersymmetric flavor-changing problem:
it is absent in theories with gauge-mediated SUSY breaking and in
theories where the scalars are kept degenerate (or aligned with the
fermions) by spontaneously broken flavor symmetries. However, there is
an important framework in a more general theory in which the
$CP$-violating signals are large. In this class of models, the dynamics
dominantly responsible for generating scalar masses is flavor-blind,
leading to nearly degenerate slepton masses.  However, the presence of
flavor-violating perturbations beneath the scale of the communication of
SUSY breaking, or of sub-dominant, flavor-violating transmission of SUSY
breaking, can give non-hierarchical, small contributions to all the
elements of the slepton mass matrices needed for a large $CP$-violating
signal. Observation of a $CP$-violating signal gives critical
information on the relationship between the nature of the communication
of SUSY breaking and the presence of new flavor-violating interactions
at high energies.

\acknowledgements

J.L.F. thanks I.~Hinchliffe for many enlightening conversations.  This
work was supported in part by the Director, Office of Energy Research,
Office of High Energy and Nuclear Physics, Division of High Energy
Physics of the U.S.  Department of Energy under Contracts
DE--AC03--76SF00098 and DE--AC02--76CH03000, and in part by the NSF
under grant PHY--95--14797.  The work of N.~A.-H. is supported by NSERC.
J.L.F. is supported by a Miller Institute Research Fellowship.

\appendix
\section{Relating $\protect\bbox{I\lowercase{m}
(\lowercase{m}^2_{12} \lowercase{m}^2_{23}
\lowercase{m}^2_{31})$ and $\widetilde{J}}$}
\label{sec:deriveJ}

In this appendix, we derive the relationship between the $CP$-violating
invariants $\widetilde{J}$ and $Im[m^2_{12} m^2_{23} m^2_{31}]$ given in
Eq.~(\ref{vanish}). Using $m^2_{\alpha \beta}= \sum_i W^*_{i \alpha}
W_{i \beta} m_i^2$, we have

\begin{equation}
Im \left[m^2_{12} m^2_{23} m^2_{31}\right] =
\sum_{ijk} Im \left[W^*_{i1} W_{i2} W^*_{j2} W_{j3} 
W^*_{k3} W_{k1}\right] m_i^2 m_j^2 m_k^2 \ .
\end{equation}
However, using the unitarity of $W$ as $W^*_{i1} W_{k1} = \delta_{ik} -
W^*_{i2} W_{k2} - W^*_{i3} W_{k3}$, the imaginary part in the above
becomes manifestly proportional to $\widetilde{J}$. Since we know that
the LHS vanishes if any two of the $m_i^2$ are identical and the RHS is
cubic in the $m_i^2$, all the $m_i^2$ dependence must be proportional to
$(m_2^2 - m_1^2)(m_3^2 - m_2^2)(m_1^2 - m_3^2)$. It is easy to show that
the constant of proportionality is 1, and Eq.~(\ref{vanish}) is
established.

\section{$\protect\bbox{\lowercase{s}}$-, 
$\protect\bbox{\lowercase{t}}$-interference}

\label{sec:stinterference}

The $CP$-violating asymmetries at colliders must be linearly dependent
on $\widetilde{J}$.  For the case of pair production, the $\sigma^{st}_0$
piece may be demonstrated to have this form by breaking the sum into 5
pieces:

\begin{eqnarray}
\sum_{i=j=k} : && \
\sum_{i} Im [ W_{i\alpha} W_{i\beta}^* 
W_{i\alpha}^* W_{i1} W_{i1}^* W_{i\beta} ] A_{ii} A_{ii} = 
0 \\
\sum_{i=j \neq k} : && \
\sum_{i\neq k} Im [ W_{i\alpha} W_{i\beta}^* 
W_{i\alpha}^* W_{i1} W_{k1}^* W_{k\beta} ] A_{ii} A_{ik} = 
\sum_{i\neq k}\left|W_{i\alpha}\right|^2
\widetilde{J}_{1\beta}^{ik}A_{ik}\\
\sum_{i=k \neq j} : && \
\sum_{i\neq j} Im [ W_{i\alpha} W_{i\beta}^* 
W_{j\alpha}^* W_{j1} W_{i1}^* W_{i\beta} ] A_{ij} A_{ii} = 
\sum_{i\neq j}\left|W_{i\beta}\right|^2
\widetilde{J}_{\alpha 1}^{ij}A_{ij}\\
\sum_{i \neq j=k} : && \
\sum_{i\neq j} Im [ W_{i\alpha} W_{i\beta}^* 
W_{j\alpha}^* W_{j1} W_{j1}^* W_{j\beta} ] A_{ij} A_{ij} = 
\sum_{i\neq j}\left|W_{j1}\right|^2
\widetilde{J}_{\alpha\beta}^{ij}A_{ij}^2\\
\sum_{i \neq j \neq k} : && \
\sum_{i \neq j \neq k} Im [ W_{i\alpha} W_{i\beta}^* 
W_{j\alpha}^* W_{j1} W_{k1}^* W_{k\beta} ] A_{ij} A_{ik} \nonumber\\
&& \ = \sum_{i \neq j \neq k} Im \left[ (\delta_{\alpha\beta} - 
W_{j\alpha} W_{j\beta}^* - W_{k\alpha} W_{k\beta}^* )
W_{j\alpha}^* W_{j1} W_{k1}^* W_{k\beta} \right] A_{ij} A_{ik} 
\nonumber\\
&& \ = \sum_{i \neq j \neq k} \left[ -
\left|W_{j\alpha}\right|^2\widetilde{J}_{1\beta}^{jk}A_{ij}A_{ik}
- \left|W_{k\beta}\right|^2
\widetilde{J}_{1\alpha}^{jk}A_{ij}A_{ik}\right] \ ,
\end{eqnarray}
where the unitarity of $W$ has been used to simplify the final term.  We
thus see that $\Delta_{\alpha\beta}^{st}$ may be written in a form
explicitly proportional to $\widetilde{J}$ as

\begin{eqnarray}
\Delta_{\alpha\beta}^{st} &=& 2 i \sum_{i\neq j} \left[
\left|W_{i\alpha}\right|^2\widetilde{J}_{1\beta}^{ij}A_{ij}
-\left|W_{i\beta}\right|^2\widetilde{J}_{1\alpha}^{ij}A_{ij}
+\left|W_{j1}\right|^2\widetilde{J}_{\alpha\beta}^{ij}A_{ij}^2 \right]
\nonumber \\
&& - 2i \sum_{i \neq j \neq k} \left[
\left|W_{j\alpha}\right|^2\widetilde{J}_{1\beta}^{jk}A_{ij}A_{ik}
- \left|W_{j\beta}\right|^2\widetilde{J}_{1\alpha}^{jk}
A_{ij}A_{ik}\right] \ .
\end{eqnarray}
The sum is over all on-shell slepton generations, and so the last sum is
relevant only when all three generations may be produced.

\begin{table}
\caption{Particle masses (in GeV) at the LHC analysis point, with 
$q = u,d,c,s$ and $l=e,\mu, \tau$.}
\begin{tabular}{cccccccc}
& Mass & & Mass & & Mass & & Mass \\ \hline
$\tilde{g}$      & 767 & 
$\tilde{b}_2$    & 662 & 
$\neutralinotwo$ & 231 &
$\tilde{l}_L$    & 239 \\

$\tilde{q}_L$    & 688 & 
$\tilde{b}_1$    & 635 & 
$\charginoone$   & 230 &
$\tilde{\nu}_l$  & 230 \\

$\tilde{q}_R$    & 662 & 
$\tilde{t}_2$    & 717 & 
$\neutralinoone$ & 121 &
$\tilde{l}_R$    & 157 \\

&&$\tilde{t}_1$  & 498 & 
&&$h$              & 100 
\end{tabular}
\label{table:masses}
\end{table}

\begin{table}
\caption{Cross sections $\sigma$ (in fb) at the LHC analysis point, with
$q = u,d,c,s$.}
\begin{tabular}{cccccc}
& $\sigma$ & & $\sigma$ & & $\sigma$ \\ \hline
$\tilde{g}\tilde{g}$ & 1751 & 
$\tilde{q}\tilde{q}$ & 2379 &
$\tilde{b}\tilde{\bar{b}}$  & 297 \\

$\tilde{g}\tilde{q},\tilde{g}\tilde{\bar{q}}$ & 8306 &
$\tilde{q}\tilde{\bar{q}}$  & 2820 &
$\tilde{t}\tilde{\bar{t}}$  & 701 
\end{tabular}
\label{table:crosssections}
\end{table}

\begin{table}
\caption{Branching fractions $B$ (in percent) at the LHC analysis point,
with $q = u,d,c,s$ and $l=e,\mu, \tau$.}
\begin{tabular}{cccccc}
& $B$ & & $B$ & & $B$ \\ \hline
$\tilde{g}\to\tilde{q}_L \bar{q}$ & 31  &
$\tilde{q}_L \to q \neutralinotwo$ & 32  &
$\neutralinotwo \to \tilde{l}_R l$ & 36  \\

$\tilde{g}\to\tilde{q}_R \bar{q}$ & 31  &
$\tilde{q}_L \to q \charginoone$ & 65  &
$\neutralinotwo \to h\neutralinoone$ & 63 \\

$\tilde{g}\to\tilde{b}_1 \bar{b}$ & 10  &
$\tilde{q}_L \to q \neutralinoone$ & 3  &
$\neutralinotwo \to Z\neutralinoone$ & 1 \\

$\tilde{g}\to\tilde{b}_2 \bar{b}$ & 14  &
$\tilde{q}_R \to q \neutralinoone$ & 99  &
$\charginoone \to W\neutralinoone$ & 98 \\

$\tilde{g}\to\tilde{t}_1 \bar{t}$ & 14  &
&&
$\tilde{l}_R \to l \neutralinoone$ & 100 
\end{tabular}
\label{table:branchingfractions}
\end{table}

\input psfig

\noindent

\noindent

\noindent

\noindent

\noindent

\noindent

\noindent

\end{document}